\documentclass[a4paper, UKenglish, cleveref, authorcolumns]{lipics-v2019}

\usepackage[utf8]{inputenc}

\usepackage{amsmath}
\usepackage{amssymb}
\usepackage{booktabs}

\usepackage{wrapfig}
\usepackage{graphicx}
\usepackage{subcaption}
\usepackage{ifthen}
\usepackage{etoolbox}
\usepackage{xspace}
\usepackage{makeidx}
\makeindex

\usepackage{tikz}
\usetikzlibrary{decorations.pathmorphing}
\tikzset{snake it/.style={decorate, decoration=snake}}
\usetikzlibrary{calc}
\usetikzlibrary{arrows}
\usetikzlibrary{shapes}
\usetikzlibrary{patterns}

\usepackage{url}

\usepackage[group-separator={,},binary-units=true,exponent-product = \cdot, output-product = \cdot]{siunitx}
\usepackage{tabularx}

\colorlet{fillColor}{blue!20}
\colorlet{fillColorAlt}{red!20}
\colorlet{fillColorStrong}{red!40}
\colorlet{frontColor}{blue}
\colorlet{frontColorAlt}{red}
\colorlet{frontColorStrong}{green}

\makeatletter

\def\concept@context{}

\newcommand{\concept@addItem}[5]{
	\expandafter\gdef\csname concept@data@#1@#2@text\endcsname{#3}%
	\expandafter\gdef\csname concept@data@#1@#2@full\endcsname{#4}%
	\expandafter\gdef\csname concept@data@#1@#2@text@pl\endcsname{#3s}%
	\expandafter\gdef\csname concept@data@#1@#2@full@pl\endcsname{#4s}%
	\expandafter\gdef\csname concept@data@#1@#2@ref\endcsname{#5}%
	\ifthenelse{\equal{#4}{}}{%
		\expandafter\gdef\csname concept@data@#1@#2@index\endcsname{#3}%
	}{
		\expandafter\gdef\csname concept@data@#1@#2@index\endcsname{#4}%
	}%
	\expandafter\gdef\csname #1#2\endcsname{%
		\@ifstar{\concept@printItem{#1}{#2}{text}{star}}{\concept@printItem{#1}{#2}{text}{}}%
	}%
  \expandafter\gdef\csname #1#2IndexAlias\endcsname##1{%
		\concept@defIndexAlias{#1}{#2}{##1}%
	}
	\expandafter\gdef\csname #1Long#2\endcsname{%
		\@ifstar{\protect{\concept@printItem{#1}{#2}{full}{star}}}{\concept@printItem{#1}{#2}{full}{}}%
	}%
	\expandafter\gdef\csname #1#2s\endcsname{%
		\@ifstar{\concept@printItem{#1}{#2}{text@pl}{star}}{\concept@printItem{#1}{#2}{text@pl}{}}%
	}%
	\expandafter\gdef\csname #1Long#2s\endcsname{%
		\@ifstar{\protect{\concept@printItem{#1}{#2}{full@pl}{star}}}{\concept@printItem{#1}{#2}{full@pl}{}}%
	}%
	\expandafter\gdef\csname #1#2Ref\endcsname{\@ifstar{%
			\concept@printItem{#1}{#2}{text}{star}%
			\concept@printReference{#1}{#2}%
		}{%
		  \concept@printItem{#1}{#2}{text}{}%
		  \concept@printReference{#1}{#2}%
	  }%
  }%

	\expandafter\gdef\csname #1Long#2Ref\endcsname{\@ifstar{%
		\concept@printItem{#1}{#2}{full}{star}%
		\concept@printReference{#1}{#2}%
	}{%
		\concept@printItem{#1}{#2}{full}{}%
		\concept@printReference{#1}{#2}%
	}%
}%

	\def\modReg{\csname concept@db@#1\endcsname}
	\ifthenelse{\equal{\modReg}{}}{%
		\expandafter\gdef\csname concept@db@#1\endcsname{#2}%
	}{%
		\expandafter\xdef\csname concept@db@#1\endcsname{\modReg,#2}%
	}%
}

\newcommand{\concept@printItem}[4]{%
	\def\itemText {\csname concept@data@#1@#2@#3\endcsname}%
	\def\itemRef  {\csname concept@data@#1@#2@ref\endcsname}%
	\def\itemIndex{\csname concept@data@#1@#2@index\endcsname}%
	\def\formatter{\csname concept@format@#1\endcsname}%
	\ifthenelse{\equal{#4}{star}}{%
		\itemText%
	}{%
		\index{\itemIndex}%
		\ifthenelse{\equal{\itemRef}{}}{%
			\formatter{\itemText}%
		}{%
			\colorlet{concept-saved-color}{.}%
			\iffalse%
			   \color{concept-saved-color}{\formatter{\itemText}}%
			\else%
			   \hyperref[\itemRef]{\color{concept-saved-color}{\formatter{\itemText}}}%
			\fi%
		}%
		\xspace%
	}%
}

\newcommand{\concept@defIndexAlias}[3]{%
		\def\itemIndex{\csname concept@data@#1@#2@index\endcsname}%
		\index{#3|see {\itemIndex}}%
}

\newcommand{\concept@printReference}[2]{%
 	\def\itemRef  {\csname concept@data@#1@#2@ref\endcsname}%
 	\ifthenelse{\equal{\itemRef}{}}{%
 		\color{red}{(No Reference specified)}%
 	}{%
	 	(see \cref{\itemRef})\xspace%
	}%
}

\newcommand{\concept@printTable}[1]{%
	\edef\modReg{\csname concept@db@#1\endcsname}%
	\def\formatter{\csname concept@format@#1\endcsname}%
	\def\tabledata{}%
	\foreach \item in \modReg {%
		\def\itemText {\csname concept@data@#1@\item @text\endcsname}%
		\def\itemFull {\csname concept@data@#1@\item @full\endcsname}%
		\def\itemRef  {\csname concept@data@#1@\item @ref\endcsname}%
		\ifthenelse{\equal{\itemRef}{}}{
			\def\itemRRef{}	
		}{
			\def\itemRRef{\autoref{\itemRef}}
		}
		\ifthenelse{\equal{\itemFull}{}}{
			\protected@xappto\tabledata{%
				\texttt{$\backslash$#1\item} & \formatter{\itemText} &---& \itemRRef \\
			}%
		}{%
			\protected@xappto\tabledata{%
				\texttt{$\backslash$#1\item}, \texttt{$\backslash$#1Long\item} & \formatter{\itemText} & \formatter{\itemFull} & \itemRRef \\
			}%
		}
	}%
	\begin{center}
	\footnotesize
	\begin{tabular}{|l|l|l|l|}\hline
		Macro & Text & Long Text & Reference\\\hline\hline
		\tabledata \hline
	\end{tabular}
	\end{center}	
}

\newcommand{\concept@indexAliases}[1]{%
	\edef\modReg{\csname concept@db@#1\endcsname}%
	\def\formatter{\csname concept@format@#1\endcsname}%
	\def\tabledata{}%
	\foreach \item in \modReg {%
		\def\itemFull {\csname concept@data@#1@\item @full\endcsname}%
		\def\itemText {\csname concept@data@#1@\item @text\endcsname}%
		\def\itemIndex{\csname concept@data@#1@\item @index\endcsname}%
		\ifthenelse{\equal{\itemFull}{}}{}{
			\index{Abbreviation!\itemText|see {\itemIndex}}
		}%
	}
}

\newcommand{\newConcept}[2]{%
	\expandafter\gdef\csname new#2\endcsname##1##2##3{%
		\concept@addItem{#1}{##1}{##2}{}{##3}%
	}%
	\expandafter\gdef\csname new#2Abbr\endcsname##1##2##3##4{%
		\concept@addItem{#1}{##1}{##2}{##3}{##4}%
	}%
	\expandafter\gdef\csname update#2Index\endcsname##1##2{%
		\expandafter\gdef\csname concept@data@#1@##1@index\endcsname{##2}%
	}%
	\expandafter\gdef\csname concept@db@#1\endcsname{}%
	\expandafter\gdef\csname tableOf#2s\endcsname{%
		\concept@printTable{#1}%
	}%
	\expandafter\gdef\csname indexAliasesOf#2s\endcsname{%
		\concept@indexAliases{#1}%
	}%
}
		
\def\concept@format@mod#1{\textsl{#1}}
\newConcept{mod}{Model}

\def\concept@format@alg#1{\textsl{#1}}
\newConcept{alg}{Algorithm}

\def\concept@format@impl#1{\textsl{#1}}
\newConcept{impl}{Implementation}

\newcommand{\Demo}[2]{\texttt{\textbackslash#1}: \colorbox{black!10}{#2{}}}
\newcommand{\demo}[1]{\Demo{#1}{\csname #1\endcsname}}
\newcommand{\demoList}[1]{{
	\def\demo@text{}
	\foreach \x [count=\i] in {#1} {%
		\ifthenelse{\equal{\i}{1}}{%
			\protected@xappto\demo@text{\demo{\x}}%
		}{%
			\protected@xappto\demo@text{, \demo{\x}}%
		}%
	}%
	\demo@text%
}}

\newcount\shortyear\newcount\shorthour\newcount\shortminute
\shorthour=\time\divide\shorthour by 60\shortyear=\shorthour
\multiply\shortyear by 60\shortminute=\time\advance\shortminute by
-\shortyear
\shortyear=\year\advance\shortyear by -1900

\def\zeit{\number\shorthour:\ifnum\shortminute<10 0\number\shortminute
	\else\number\shortminute\fi}

\def\rightmark{{\small \today, \zeit{}}}

\makeatother %
\newModelAbbr{BA}{BA}{Barab\'asi-Albert}{subsec:mod_preferential_attachment}
\newModelAbbr{BTER}{BTER}{Block Two-Level Erd\H{o}s-R\'enyi}{subsec:mod_bter}
\newModelAbbr{GBTER}{GBTER}{Generalized BTER}{subsec:mod_bter}
\newModelAbbr{EGBTER}{EGBTER}{Extended GBTER}{subsec:mod_bter}
\newModelAbbr{ABTER}{A-BTER}{Adapted BTER}{subsec:mod_bter}
\newModelAbbr{CB}{CB}{Curveball}{subsec:rand_curveball}
\newModelAbbr{ChungLu}{CL}{Chung-Lu}{subsec:mod_chung_lu}
\newModelAbbr{CM}{CM}{Configuration Model}{subsec:rand_configuration_model}
\newModelAbbr{ECM}{ECM}{Erased Configuration Model}{subsec:rand_configuration_model}
\newModelAbbr{ES}{ES}{Edge-Switching}{subsec:rand_edge_switching}
\newModelAbbr{ER}{ER}{Erd\H{o}s-R\'enyi}{subsec:mod_erdos_renyi}
\newModelAbbr{GCB}{G-CB}{Global Curveball}{subsec:rand_curveball}
\newModelAbbr{RGG}{RGG}{Random Geometric Graph}{subsec:mod_geometric_graphs}
\newModelAbbr{RDT}{RDT}{Random Delaunay Triangulation}{subsec:mod_delaunay_triangulation}
\newModelAbbr{RHG}{RHG}{Random Hyperbolic Graph}{subsec:mod_hyperbolic_graphs}
\newModelAbbr{GIRG}{GIRG}{Geometric Inhomogenous Random Graphs}{subsec:mod_hyperbolic_graphs}
\newModel{Gilbert}{Gilbert}{subsec:mod_gilbert}
\newModel{Kronecker}  {Kronecker Graph}{subsec:mod_rmat}
\newModel{StocKronecker}  {Stochastic Kronecker Graph}{subsec:mod_rmat}
\newModelAbbr{RMAT}{R-MAT}{Recursive Matrix Model}{subsec:mod_rmat}
\newModelAbbr{SBM}  {SBM}{Stochastic Block Model}{subsec:mod_sbm}
\newModel{PPM}{Planted Partition Model}{subsec:mod_sbm}
\newModel{IRG}{Inhomogenous Random Graph}{subsec:mod_sbm}
\newModelAbbr{FDSM}{FDSM}{Fixed-Degree-Sequence-Model}{sec:graph-randomization}
\newModel{LFR}  {LFR}{subsec:mod_lfr}
\newModel{CKB}  {CKB}{subsec:mod_ckb}
\newModelAbbr{AGM}{AGM}{Community-Affiliation Graph Model}{subsec:mod_ckb}
\newModel{Darwini}  {Darwini}{subsec:mod_darwini}
\newModel{dKGraphs}  {$dK$-graphs}{subsec:mod_dkgraphs}
\newModel{NodeCopy}{Node Copy}{subsubsec:mod_copy_model}
\newModelAbbr{WRG}{WRG}{Weighted Random Graph}{subsec:special_graphs_weighted}
\newModelAbbr{WS}{WS}{Watts-Strogatz}{}
\newModelAbbr{ERGM}{Exponential-Family Random Graph Models}{ERGMs}{sec:future-challanges}

\newModel{ReCon}{ReCon}{}
\newModel{MUSKETEER}{MUSKETEER}{subsec:graph_repl_multiscale}
\newModel{SpectralGen}{SpectralGen}{subsec:graph_repl_multiscale}

\updateModelIndex{AGM}{Community-Affil. Graph Model}
\updateModelIndex{BA}{Barab\'asi-Albert Model}
\updateModelIndex{ChungLu}{Chung-Lu Model}
\updateModelIndex{ER}{Erd\H{o}s-R\'enyi's $G(n, m)$ Model}
\updateModelIndex{Gilbert}{Gilbert's $G(n, p)$ Model}
\updateModelIndex{WS}{Watts-Strogatz Model}
\updateModelIndex{GIRG}{Geometric Inhomo. Random Graphs}
\updateModelIndex{FDSM}{Fixed Degree Seq. Model}

\newAlgorithm{HyperGen}{HyperGen}{subsubsec:mod_hyperbolic_graphs_threshold}
\newAlgorithm{RHG}{RHG}{subsubsec:mod_hyperbolic_graphs_threshold}
\newAlgorithm{sRHG}{sRHG}{subsubsec:mod_hyperbolic_graphs_threshold}
\newAlgorithm{NkBand}{NkBand}{subsubsec:mod_hyperbolic_graphs_threshold}
\newAlgorithm{NkQuad}{NkQuad}{subsubsec:mod_hyperbolic_graphs_general}
\newAlgorithm{NkBandHT}{NkBandHT}{subsubsec:mod_hyperbolic_graphs_general}
\newAlgorithm{GIRGs}{GIRGs}{subsec:mod_girg}
\newAlgorithm{HyperGIRGs}{HyperGIRGs}{subsec:mod_girg}
\newAlgorithm{HavelHakimi}{Havel-Hakimi}{sec:graph-randomization}
\newAlgorithm{KronFit}{KronFit}{subsec:mod_rmat}
\newAlgorithm{Deg}{DEG}{subsec:special_graphs_regular}
\newAlgorithm{IncReg}{INC-REG}{subsec:special_graphs_regular}
\newAlgorithm{FisherYatesShuffle}{Fisher-Yates shuffle}{subsec:prelim-random-permutation}
\newAlgorithm{KnuthShuffle}{Knuth shuffle}{subsec:prelim-random-permutation}

\newImplementation{KaGen}{KaGen}{}
\newImplementation{NetworKit}{NetworKit}{}

\DeclareMathOperator{\sort}{sort}
\DeclareMathOperator{\scan}{scan}

\newcommand{\symOh}{\ensuremath{\mathcal{O}}}

\newcommand{\Oh}[1]{\ensuremath{\symOh\!\left(#1\right)}}
\newcommand{\Ohsmall}[1]{\ensuremath{\symOh(#1)}}

\newcommand{\ie}{i.\,e.,\xspace}
\newcommand{\eg}{e.\,g.,\xspace}
\newcommand{\etal}{et~al.\xspace}

\newcommand{\wrt}{w.\,r.\,t.\xspace}
\newcommand{\cf}{cf.\xspace}

\newcommand{\nproc}{\ensuremath{P}}

\newcommand{\opstyle}[1]{\mathrm{%
		{#1}%
}}
\newcommand{\degree}{\ensuremath{\opstyle{deg}}}

\newcommand{\maxdeg}{\ensuremath{\opstyle{maxdeg}}}

\newcommand{\dist}{\ensuremath{\opstyle{dist}}}
\newcommand{\adj}{\ensuremath{\opstyle{N}}}

\newcommand{\degseq}{\ensuremath{\mathcal{D}}}
\newcommand{\cc}{\ensuremath{\opstyle{cc}}}
\newcommand{\density}{\ensuremath{\opstyle{dens}}}

\newcommand{\gGn}{\ensuremath{\mathbb G(n)}\index{$\mathbb G(n)$!set}\xspace} %
\newcommand{\gGnm}{\ensuremath{\mathbb G(n,m)}\index{$\mathbb G(n,m)$!set}\xspace} %
\newcommand{\gGnr}{\ensuremath{\mathbb G^{(r)}(n)}\xspace} %
\newcommand{\Gn}{\ensuremath{\mathcal{G}(n)}\index{$G(n)$!uniform distr.}\xspace} %
\newcommand{\Gnm}{\ensuremath{\mathcal{G}(n,m)}\index{$G(n,m)$!uniform distr.}\xspace} %
\newcommand{\Gnp}{\ensuremath{\mathcal{G}(n,p)}\index{$\mathcal G(n, p)$}\xspace}
\newcommand{\Gnr}{\ensuremath{\mathcal{G}^{(r)}(n)}\xspace} %

\newcommand{\sNpos}{\ensuremath{\mathbb{N}_{>0}}}

\newcommand{\sR}{\ensuremath{\mathbb{R}}}

\newcommand{\symprob}{\ensuremath{\mathbb{P}}}

\newcommand{\prob}[1]{\ensuremath{\symprob\!\left[#1\right]}}

\newcommand{\geom}[1]{\ensuremath{\opstyle{Geom}(#1)}}
\newcommand{\whp}{whp.\xspace}

\newcommand{\Def}{\ensuremath{:=}}
\newcommand{\seq}[3]{\ensuremath{[\,#1\,]_{#2}^{#3}}}

\newcommand{\avg}[1]{\ensuremath{\left\langle #1 \right\rangle}}
\newcommand{\avgdeg}{\ensuremath{\avg{\degree}}}
\newcommand{\avgdist}{\ensuremath{\avg{\dist}}}

\newcommand{\floor}[1]{\ensuremath{\left\lfloor #1\right\rfloor}}
\newcommand{\card}[1]{\ensuremath{\lvert #1\rvert}}

\newcommand{\showcomment}[1]{#1}
\renewcommand{\showcomment}[1]{} %

\def\indConvertEM#1#2\relax{%
   \ifx#1!$\to$\else#1\fi%
   \ifx#2\relax\else\indConvertEM#2\relax\fi%
}

\newcommand{\newindex}[2]{%
   \expandafter\gdef\csname ind#1\endcsname{\index{#2}}%
   \expandafter\gdef\csname indDef#1\endcsname{\index{#2|textbf}}%
   \expandafter\gdef\csname indText#1\endcsname{\indConvertEM#2\relax\relax}%
   \expandafter\gdef\csname indBegin#1\endcsname{\index{#2|(}}%
   \expandafter\gdef\csname indEnd#1\endcsname{\index{#2|)}}%
}

\newindex{RejectionSampling}{Sampling!Rejection}
\index{Rejection sampling|see \indTextRejectionSampling}
\newindex{BernoulliSampling}{Sampling!Bernoulli}
\index{Bernoulli sampling|see \indTextBernoulliSampling}
\newindex{KoutOfNSampling}{Sampling!Hypergeometric}
\index{Sampling!$k$ out of $[N]$|see \indTextKoutOfNSampling}
\newindex{WeightedSampling}{Sampling!Weighted}
\index{Weighted sampling|see  \indTextWeightedSampling}
\newindex{RandomPermutation}{Random Permutation}

\newindex{DistributedMemory}{Distributed Memory}
\newindex{CommunicationFree}{Communication-free}
\newindex{IOEfficient}{I/O-efficient}
\index{External memory|see \indTextIOEfficient}
\newindex{Streaming}{Streaming}

\newindex{GPU}{GPU}

\newindex{GraphPlanar}{Graph!Planar}
\index{Planar Graph|see \indTextGraphPlanar}
\newindex{GraphDirected}{Graph!Directed}
\index{Directed Graph|see \indTextGraphDirected}
\newindex{GraphWeighted}{Graph!Weighted}
\index{Weighted Graph|see \indTextGraphWeighted}
\newindex{GraphSpatial}{Graph!Spatial}
\index{Spatial Graph|see \indTextGraphSpatial}
\newindex{GraphRegular}{Graph!Regular}
\index{Regular Graph|see \indTextGraphRegular}
\newindex{GraphDensity}{Graph!Density}
\index{Density|see \indTextGraphDensity}

\newindex{DelaunayTriangulation}{Delaunay triangulation}
\newindex{PreferentialAttachment}{Preferential attachment}
\newindex{QuadTree}{Quad tree}

\newindex{DegreeSequence}{Degree sequence}
\newindex{DegreeSequenceGraphical}{\indTextDegreeSequence!Graphical}
\newindex{DegreeSequenceRegular}{\indTextDegreeSequence!Regular}

\newindex{SelfLoops}{Graph!Self loops}

\newindex{PowerLaw}{Power law distribution}
\newindex{Clustering}{Graph!Clustering}
\newindex{Diameter}{Graph!Diameter}

\newindex{CommunityDetection}{Community Detection!Benchmark}
\newindex{PlantedCommunity}{Community Detection!Planted Communities}

\title{Recent Advances in Scalable Network Generation}

\authorrunning{M. Penschuck et al.} %
\Copyright{Manuel Penschuck and Ulrik Brandes and Michael Hamann and Sebastian Lamm and Ulrich Meyer and Ilya Safro and Peter Sanders and Christian Schulz}
\acknowledgements{%
We would like to thank the organizers and the staff of the Dagstuhl seminar~18241 on ``High-Performance Graph Algorithms'' during which the initial idea for this survey was formed.
As usual in algorithmics the order of authors is alphabetical with the exception that we moved Manuel Penschuck to the front since he coordinated this project, and contributed a more than proportional amount of material. %
}

\author{Manuel Penschuck}{Goethe University Frankfurt}{penschuck@algorithm.engineering}{}{}
\author{Ulrik Brandes}{ETH Zürich}{ubrandes@ethz.ch}{}{}
\author{Michael Hamann}{Karlsruhe Institute of Technology}{michael.hamann@kit.edu}{}{}
\author{Sebastian Lamm}{Karlsruhe Institute of Technology}{sebastian.lamm@kit.edu}{}{}
\author{Ulrich Meyer}{Goethe University Frankfurt}{umeyer@ae.cs.uni-frankfurt.de}{}{}
\author{Ilya Safro}{Clemson University}{isafro@clemson.edu}{}{}
\author{Peter Sanders}{Karlsruhe Institute of Technology}{sanders@kit.edu}{}{}
\author{Christian Schulz}{University of Vienna}{christian.schulz@univie.ac.at}{}{}
\titlerunning{Recent Advances in Scalable Network Generation}

\funding{%
   This work was partially supported by Deutsche Forschungsgemeinschaft (DFG) under grants
   BR~2158/11-2,
   ME~2088/3-2,
   ME~2088/4-2, 
   SCHU~257/1-2,
   and WA~654/22-2.
   The research leading to these results has received funding from the European Research Council under the European Community's Seventh Framework Programme (FP7/2007-2013) / ERC grant agreement No. 340506.
}

\hideLIPIcs
\nolinenumbers

\makeatletter%
\begin{document}
   \maketitle
   \begin{abstract}
Random graph models are frequently used as a controllable and versatile data source for experimental campaigns in various research fields.
Generating such data-sets at scale is a non-trivial task as it requires design decisions typically spanning multiple areas of expertise.
Challenges begin with the identification of relevant domain-specific network features,
 continue with the question of how to compile such features into a tractable model,
  and culminate in algorithmic details arising while implementing the pertaining model.

In the present survey, we explore crucial aspects of random graph models with known scalable generators.
We begin by briefly introducing network features considered by such models, and then discuss random graphs alongside with generation algorithms.
Our focus lies on modelling techniques and algorithmic primitives that have proven successful in obtaining massive graphs.
We consider concepts and graph models for various domains (such as social network, infrastructure, ecology, and numerical simulations),
 and discuss generators for different models of computation (including shared-memory parallelism, massive-parallel GPUs, and distributed systems).    \end{abstract}
   \clearpage

   %

\section{Introduction}
\label{sec:intro}
Generating synthetic networks is one of the most active research areas in network science and general graph algorithms.
Theoreticians, domain experts, and algorithm developers need high-quality network data for various purposes such as algorithm engineering, decision-making, and simulations.
However, obtaining real-world network data is often accompanied with various obstacles.
Some of the reasons for the resulting shortage include classified or proprietary information, legal hindrances, and economic considerations to obtain high-quality data as well as simple lack of future data in evolving networks.
For very large graphs processed on supercomputers, transferring the graphs to data centers, storing them indefinitely on disks, and loading or distributing them for experiments can also be quite expensive.
In these cases, synthetic networks generated by models are used to substitute the real-world networks.

There are two major considerations in the process of designing a synthetic network generator.
Firstly, generating data from a high-quality synthetic model requires reproducing important structural properties observed in reference data.
Alternatively, one may consider to generate networks with predefined properties but with no original reference (or only much smaller examples). %
Each case requires an unbiased coverage of the entire set of relevant networks.
Depending on the complexity of the structural properties, a generator may need to solve optimization problems of varying complexity.
Secondly, for many applications, the generation process must be sufficiently scalable to produce large-scale instances in a reasonable time with respect to the available computational resources.
For example, there is little practical value in generating a small network of several tens of nodes for social information network analysis.

These considerations suggest a traditional quality/runtime trade-off that must be taken into account when designing or using generators for large-scale networks.
In particular, this is important when many large-scale synthetic networks are required for computational experiments;
\eg in order to demonstrate the robustness of an algorithm, to gather sufficient statistical information, or to observe the long-term evolution of a network.
While there exist several comprehensive surveys~\cite{survey:rg,goldenberg2010survey,bonato2004survey}
that focus on the quality component of this trade-off, we find that not many pay close attention on the combination.
In this survey, we discuss various classes of generation methods.
We focus on their scalability, algorithmic, and implementation aspects, as well as on the relevance of different parallel programming models to maintain acceptable performance in the generation of large-scale networks.
 
\section{Graph properties and uses of generators}
\label{sec:graphprop}
The set $\gGn$ of all simple undirected graphs with~$n$ vertices contains $2^{\binom{n}{2}}$~graphs.
We define the model $\Gn$ as the uniform distribution over $\gGn$.
The $\Gn$ model has proven useful for the probabilistic method in existential combinatorics~\cite{bollobas1985random},
but it is not at all a plausible assumption for statistics of empirical graphs.
Just consider that a graph sampled uniformly at random from $\gGn$
contains half of the $\binom{n}{2}$ edges in expectation---which is much more than typically found in empirical networks (see below).

To better match empirical distributions of graphs encountered in different domains of application, models are devised in which certain graph invariants can be controlled for.
In this section, we provide background on some of such properties and a number of important model classes.
The section concludes with an outline of a few ways in which generators parameterized on graph properties are commonly applied.
 
\subsection{Graph properties}
\label{subsec:graph-properties}
We next recall definitions of a number of graph properties
that are commonly used to constrain or skew
the distribution of graphs generated from a model.
All models referred to below are discussed in more detail in later sections.

\begin{table}
	\caption{Notation used in this survey}
	\label{tab:notation}
	\begin{center}
	\small
	\iftrue\def\optLineBreak{\\ & and }\else\def\optLineBreak{}\fi
	\begin{tabular}{l l}\hline
		$\seq{x_i}{i=a}{b}$, $[a .. b]$, $[k]$ & Sequence or set $[x_a, x_{a+1}, \ldots, x_b]$, $[a .. b] \Def \seq{i}{i=a}{b}$, $[k] \Def [1..k]$ \\\hline
		$\avg{X}$ & Average, $\avg{X} \Def \frac 1 n \sum_{v \in V} X(v)$ \\\hline
		$\cc(v)$, $\cc(G)$ & Clustering coefficient $\cc(v) \Def \density(G[\adj(v)])$,  $\cc(G) \Def \avg{\cc}$\\\hline		
		$\density(G)$ & Density, $\density(G) \Def m/\binom{n}{2}$ if $n > 1$ else $\density(G) = 0$ \\\hline		
		$\dist(u, v)$ & Length (number of edges) of shortest path between~$u$ and~$v$ \\\hline
		$G = (V, E)$ & Graph with nodes~$V = [n]$ and edges~$E\subseteq\binom{V}{2}$ \\\hline
		$G[V']$ & Induced subgraph~$(V', E')$ if $G {=} (V, E)$, s.t. $V' \subseteq V$, \optLineBreak $E' = \{e\, |\, e {=} \{u{,}v\} {\in} E\, \land\, u{,} v {\in} V'\}$ \\\hline
		$\gGn$, $\gGnm$ & Set of graphs with $\card{V}=n$ and $\card{E} =m$, respectively \\\hline
		$\Gn$, $\Gnm$ & Uniform distributions over $\gGn$ and $\gGnm$, respectively \\\hline
		$n$, $m$ & Number~$n$ of nodes, number~$m$ of edges \\\hline
		$\adj(v)$, $\degree(v)$ & Neighbors $\adj(v) \Def \{ u : \{u, v\} \in E \}$, degree $\degree(v) \Def |\adj(v)|$ \\\hline
		\whp & With high probability: true with probability at least $1 - 1/n^c$ for $c \ge 1$\\\hline
	\end{tabular}
	\end{center}
\end{table}

\subsubsection*{Density}
\indDefGraphDensity
We have parameterized the class $\gGn$ of all graphs
by their \emph{order}~$n$, \ie the number of vertices.
The other obvious parameter to control for is~$m$, the number of edges.
The overwhelming majority of graphs in $\gGn$ is dense,
\ie\ $m=\Theta(n^2)$,
but graphs arising in application domains are often \emph{sparse},
say $m = \Oh{n\log n}$.
This may be because, for instance, edges are associated with costs
or vertices have limited capacity to be adjacent with others.

For any given combination of~$n$ and $0\leq m\leq\binom{n}{2}$,
we obtain a subclass $\gGnm\subseteq\gGn$ of graphs $G=(V,E)$ with $V=[n]$ and $\card{E}=m$.
They all have the same %
\emph{density}~$\density(G) \Def m/\binom{n}{2}$,
where we define $\density(G) \Def 0$ for $n \in \{0, 1\}$.
Since the sum of degrees equals twice the number of edges,
all graphs in $\gGnm$ also have the same
\emph{average degree}~$\avgdeg \Def \frac{1}{n}\sum_{v\in V}\degree(v)= 2m / n$.

Analogously to $\Gn$, we denote the uniform distribution on $\gGnm$ by $\Gnm$.
While $\Gnm$ was the original model of Erd{\H o}s and R\'enyi~\cite{erdds1959random},
Gilbert~\cite{gilbert1959random} introduced the related $\Gnp$ model defined on all of $\gGn$ where each edge is present independently with probability~$p$ (see \cref{sec:erdos_related}).
Thus, the number of edges, and as a consequence density and average degree, are fixed only in expectation.
Note that $\Gnp=\Gn$ for $p=1/2$.

Random graphs in each of the three models are \emph{balanced},
in the sense that the expected density of any vertex-induced subgraph is constant.
\indDefClustering%
To assess differences in local density of a graph $G=(V,E)$, the \emph{clustering coefficient} of a vertex~$v\in V$ is defined as $\cc(v) \Def \density(G[\adj(v)])$, \ie the density of the subgraph induced by $v$'s neighborhood.
The average over all vertices is called
the clustering coefficient $\cc(G) \Def \frac{1}{n}\sum_{v\in V}\cc(v)$ of the graph.

\indGraphSpatial%
Models that favor local cohesion (\ie produce substructures with an above-average internal density) include \emph{random geometric graphs} (see \cref{sec:spatial_graphs}).
Here, vertices are randomly assigned to positions, \eg in a Euclidean or hyperbolic space, and edge probabilities are made dependent on the distance in that space.

Heterogeneity in local density is introduced in
\emph{planted partition} or \emph{stochastic blockmodels} (see \cref{sec:block_models}),
where vertices are partitioned into subsets,
and different edge probabilities are assigned
depending on which subsets the endpoints are in.

\subsubsection*{Degrees}
\indDefDegreeSequence%
Similar to density, there may be a tendency for degrees
to be distributed unevenly in a graph.
The subset $\gGnm$ can be restricted further
by fixing not only the number of edges
but the entire \emph{degree sequence}, \ie using a sequence~$\degseq \Def \seq{d_i}{i=1}{n}$ to prescribe the degree~$\degree(v) = d_v$ of each node~$v\in[n]$.
This is done in \emph{fixed degree sequence models} as described in \cref{sec:graph-randomization}.

As was done for density in the case of $\Gnp$,
we may also want to sample from all of $\gGn$ with degrees fixed only in expectation.
One such model is introduced in Chung and Lu~\cite{chung2002connected}
and uses vertex weights to construct a rank-1 matrix of edge probabilities (see \cref{subsec:mod_chung_lu}).
In an even more relaxed setting, the \emph{degree distribution} (\ie the relative frequencies of degrees)
is prescribed by a closed-form function, the shape of which can be controlled via a small number of parameters.
The most notable example of this kind are \emph{scale-free} degree distributions,
where the share of vertices of degree~$k$
is approximately proportional to $k^{-\gamma}$ for some constant~$\gamma>1$.
\indPreferentialAttachment%
These are obtained, for instance, from \emph{preferential attachment models} (see \cref{subsec:mod_preferential_attachment}).

\subsubsection*{Distances}

Densities and degrees are properties that involve counts and frequencies of edges, and therefore rather straightfoward to integrate in the specification of random graph models. Prime examples of properties that give rise to more complicated structural dependencies relate to distances.

The \emph{(shortest-path or graph-theoretic) distance}, $\dist(s,t)$,
between two vertices $s,t\in V$ is the length of a shortest $(s,t)$-path,
\ie the minimum number of edges in any path connecting them.
The distance is defined to be infinite,
if the two vertices are in different connected components.
In the remainder of this section we assume graphs to be connected to avoid the treatment of special cases.
Some remarks on connectivity are made in \cref{subsec:special_graphs_connected}.

Analogously to the average degree, the \emph{average distance} (or \emph{characteristic path length})
$\avgdist=\sum_{s,t\in V} \dist(s,t) /\binom{n}{2}$
is used as a criterion to discriminate certain graph structures.
When representing, for example, social relations,
the resulting graphs typically have
bounded average degree, $\avgdeg = \Oh{1}$
and small average distance, $\avgdist = \Oh{\log n}$.
These features, together with a high clustering coefficient,
are characteristic of \emph{small-world models}.

In contrast, the random spatial graphs discussed in \cref{subsec:mod_geometric_graphs} have large average distance (typically $\Oh{n^{1/d}}$ for $d$-dimensional Euclidean spaces).
It is however possible to control the average distance by considering different aspect ratios of the boxes used for allocating points, or by varying the underlying geometry alltogether.

\indDefDiameter%
More strict than average distance is the \emph{diameter} of a graph,
defined as the maximum distance between any two of its vertices.
There are multiple ways to define \emph{distance sequences}~\cite{buckley1990distance}
but, unlike degree sequences,
none is commonly used in the definition of random graph models.

\subsubsection*{Graph classes}
Other graph properties are not expressed via aggregates of lower-level indices at all.
\indGraphPlanar%
For instance, a graph is called \emph{planar},
if it can be drawn in the plane without edge crossings.
While, as a consequence of this criterion, they are sparse ($m\leq 3n-6$),
there is no known parametrization
in terms of the number of elements, density, degrees, or distances,
that could be used for random planar graph models (see \cref{subsec:mod_planar}).

On the other hand, a \emph{split graph} is a graph that has a partition into
an induced clique (a complete subgraph)
and an induced independent set (an empty subgraph).
Split graphs are an idealized version
of what is called a \emph{core-periphery structure}~\cite{borgatti2000coreperiphery}.
Despite this definition in global terms,
split graphs can indeed be characterized using degree sequences only~\cite{hammer1981split}.

For each class of graphs,
a corresponding random graph model can be defined
by the uniform distribution on its elements.
Since generators for such models generally depend on specific class characteristics,
we consider only random planar graphs
as a particularly interesting and important example
of a class that does not lend itself easily to modeling.
 
\subsection{Use cases}
\label{subsec:overview-applications}
The most direct and, at least in computer science, most common use of graph generators is in the creation of input instances for computational experiments~\cite{mcgeoch2012experimental}.
Such experiments typically serve to establish response curves recording levels of an outcome (dependent) variable as function of independent variables.
Especially in the context of scaling experiments, independent variables are often related to the graph size (\eg number of nodes, number of edges, average degree), or specific model parameters.
Common outcome variables include implementation performance (\eg running time or solution quality), process characteristics (\eg spreading of information, resilience against attack), or other graph invariants (\eg connectivity, graph spectra).

Covering a desired experimental region with benchmark data is often impractical or even unrealistic.
To control experimental factors, it is necessary that instances in a computational experiment satisfy certain constraints or show tendencies with respect to certain properties.
Empirical data of sufficient coverage and variability may not be available, samples may not be sufficiently representative, or the experimental region may be too vast to store the instances explicitly.
In such cases, generative models are a convenient means to fill the void~\cite{DBLP:conf/beliv/SchulzNEFHHKNSB16}.

In the same way that the parameters of $\Gnp$ allow to control the order and density of graphs, other models are used to generate instances that vary along a number of dimensions while ensuring a certain distribution, or the presence or absence of certain other properties.
In addition to systematic variation of parameters when sampling from a distribution or construction process, there are a number of techniques for graph generation that can be used to augment a given set of benchmark data.
These include sampling from models with some parameters learned from the benchmark data~\cite{brandes2011offline}, perturbation or systematic variation of given instances~\cite{borgatti2006robustness}, and scaling these instances by creating larger or smaller graphs with similar structure (see \cref{sec:graph_replication}).

We note that the randomized generation of instances from an explicit or implicit distribution is also a technique in computer graphics and the arts~\cite{devroye1995botanical}.
In algorithmic art it has been suggested that a generated instance should be viewed as representing both itself and the ensemble from which it was sampled~\cite{nake2006doppelt}.
This artistic view is indeed related to computational model-based inference where a focal instance is compared to an ensemble of generated instances with the goal to establish whether the given instance exhibits specific features to a degree that is common or unusual for the sample~\cite{snijders2011models}.
A common ensemble is generated from the conditionally uniform model that assigns equal probability to all graphs with the same degree sequence (see \cref{sec:graph-randomization}).

A very recent application is the randomized design of neural network architectures~\cite{xie2019iccv}.
 
\section{General Algorithmic models and techniques}\label{sec:algo-techniques}

\subsection{Models of computation}
\label{subsec:prelim_compmods}
The scalability of a network generation algorithm is connected to the assumed model of computation.
In particular, we want algorithms that run in parallel on~$\nproc$ identical processors, and require work close to the that of a good sequential generator---at least when the generated data-sets are large.
In this survey, we discuss the algorithms at an abstract level, and consider whether and how they are implementable on different parallel architectures such as GPUs, shared- and distributed-memory.

\indDefGPU%
\indDefDistributedMemory%
For distributed-memory models, an important aspect are communication costs which easily dominate the overall costs when large data-sets are processed.
An algorithm is \emph{communication efficient} if the communication volume per processor is sublinear in the required local work~\cite{SSM13}.
For graph generators, we can even achieve (essentially) \emph{communication-free} algorithms~\cite{DBLP:conf/ipps/FunkeL00SL18}.
See the discussion in \cref{subsec:mod_geometric_graphs} for an example how to make a parallel graph generator communication-free.
\indDefCommunicationFree%

Another aspect of scalability is that large graphs may not fit into the main memory of the machine.
\indDefIOEfficient%
Hence, it makes sense to consider \emph{external memory algorithms}~\cite{AggarwalVitter88} where fast random access is limited to a bounded internal memory of size~$M$, and the external memory is accessed in blocks of size~$B$.
An important special case are \emph{streaming algorithms} for graph generation that output edges one at a time without requiring access to the entire graph.
\indDefStreaming%
 	\subsection{Random Permutations}
	\label{subsec:prelim-random-permutation}
\indDefRandomPermutation%
The \algFisherYatesShuffle (or \algKnuthShuffle)~\cite{Knu81} obtains a random permutation of an array $A[1 .. n]$ in time $\Oh{n}$.
Conceptually, it places all items into an urn, draws them sequentially without replacement, and returns the order in which they were drawn.
The algorithm works in-place, and fixes the value of $A[i]$ in iteration~$i$ by swapping $A[i]$ with $A[j]$ where~$j$ is chosen uniformly at random from the not yet fixed positions $[i..n]$.
A random permutation can be computed in parallel by~$P$ processors by assigning each element to one of~$P$ buckets uniformly at random and then applying the sequential algorithm to each bucket~\cite{San98c}.
\indIOEfficient%
A similar technique yields an I/O-efficient random permutation algorithm~\cite{San98c}.

	\subsection{Basic sampling techniques}
	\label{subsec:prelim-basic-sampling-tech}
In this section we discuss sampling primitives, \eg how to generate random numbers with an underlying distribution or how to uniformly take~$n$ elements from a universe.

\iftrue
\begin{figure}
	\begin{center}
\begin{tikzpicture}
	\def\sizex{0.68}
	\def\sizey{0.68}

	\fill[fill=fillColor] (1* \sizex, 0) to ++ (1* \sizex, 0) to ++ (0, \sizey) to ++ (-1* \sizex, 0);
	\fill[fill=fillColor] (6* \sizex, 0) to ++ (1* \sizex, 0) to ++ (0, \sizey) to ++ (-1* \sizex, 0);

	\foreach \x in {1, ..., 7} {
		\path[draw] (\x * \sizex, 0) to (\x * \sizex, \sizey);
	}

	\path[draw] (0, 0) to  ++ (8* \sizex, 0);
	\path[draw] (0, \sizey) to ++ (8* \sizex, 0);

	\path[draw, snake it, segment length=5] (0, 0) to ++ (0, \sizey);
	\path[draw, snake it, segment length=5] (8* \sizex, 0) to ++ (0, \sizey);
	
	\node (sampled-label1) at (0.8* \sizex, 1.4) {sampled};
	\path[draw, bend right, ->] (sampled-label1) to (1.5* \sizex, 0.5);

	\node (sampled-label2) at (7.2* \sizex, 1.4) {sampled};
	\path[draw, bend left, ->] (sampled-label2) to (6.5* \sizex, 0.5);

	\path[draw, thick] (2.3* \sizex, 0.5) to ++(0, 1.3);
	\path[draw, thick] (5.7* \sizex, 0.5) to ++(0, 1.3);
	\path[draw, thick, <->] (2.4* \sizex, 1.8) to
		node[above] {skip distance $S$}
		node[below, align=center, font=\small] {$\prob{S = k}$ \\ $= (1-p)^k p$}
		 ++(3.2* \sizex, 0);
\end{tikzpicture} 	\end{center}

	\caption{%
		Bernoulli sampling by skipping. Output-sensitive Bernoulli sampling with probability~$p$ by skipping~$S$ items where $S{+}1$ follows the geometric distribution $\geom{p}$.
	}
	\label{fig:bernoulli-skip}
\end{figure}
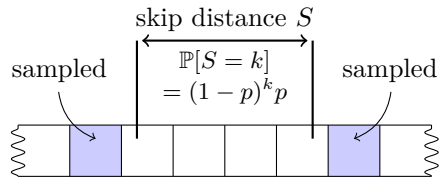
\fi

Many of the standard distributions can be sampled in constant (expected) time with well known techniques (\eg~\cite{DBLP:books/sp/Devroye86, PreEtAl92e}).
In this survey we need geometric, binomial, and hypergeometric random deviates~\cite{Stad90hyp}. %
Respective generators are often arithmetically expensive since they require the evaluation of transcendental functions.
When many deviates with known parameters have to be computed, this can be greatly accelerated by vectorization taking advantage of SIMD instructions or GPUs~\cite{DBLP:journals/toms/SandersLHSD18}.
Often, software libraries doing this are already available.

\subsubsection{Bernoulli sampling}\label{sss:BernoulliSampling}
\indDefBernoulliSampling%
Sampling each element from $[N]$ with probability~$p$ can be done directly in time $\Oh{N}$ by throwing~$N$ coins that show head with probability $p$.
As illustrated in \Cref{fig:bernoulli-skip}, it can be made to work in time proportional to the output size by generating distances between sampled elements which have a geometric distribution~\cite{10.2307/2281647} with parameter $1/p$. Bernoulli sampling is easy to parallelize and vectorize since all samples are independent.

\subsubsection[Sampling k elements from {[n]}]{Sampling~$k$ elements from {$[N]$}}\label{sss:uniformSampling}
\iftrue
\begin{figure}
	\begin{center}
\begin{tikzpicture}
	[index label/.style={anchor=north, inner sep=0.1em, font=\footnotesize}]
	\def\boxSize{0.3}
	\newcommand{\fmtSet}[1]{\textcolor{frontColor}{#1}}
	\newcommand{\fmtCnt}[1]{\textcolor{frontColorAlt}{#1}}
		
	\def\drawArray#1#2#3#4{
		\def\xx{#1}\def\yy{#2}\def\n{#3}\def\sampled{#4}

		\path[draw] (\xx, \yy) to ++(\n * \boxSize,0);
		\path[draw] (\xx, \yy + \boxSize) to ++(\n * \boxSize,0);

		\foreach \x in \sampled {
			\filldraw[fill=fillColor] (\xx + \x * \boxSize, \yy)
			to ++(\boxSize, 0)  to ++(0, \boxSize) to ++(-\boxSize, 0);
		}		

		\foreach \x in {0, ..., \n} {
			\path[draw]	(\xx + \x * \boxSize, \yy) to ++ (0, \boxSize);
		}
	}

	\def\upperY{1.25}
	\drawArray{0}{\upperY}{24}{1,3,4,8,14,23}
	\node[anchor=south] at (12 * \boxSize, \upperY + \boxSize) {Sample \fmtCnt{$k$} items from \fmtSet{$[N]$}};
	\path[draw, very thick] (12 * \boxSize, \upperY - \boxSize / 2) to ++ (0, 2*\boxSize);

	\node[anchor=south, inner sep=0, font=\small] at (3*\boxSize, 0.3) {\fmtCnt{$X$} items from \fmtSet{$[1 .. N']$}};
	\drawArray{-3*\boxSize}{0}{12}{1,3,4,8}

	\node[anchor=south, inner sep=0, font=\small] at (21*\boxSize, 0.3) {\fmtCnt{$k{-}X$} items from \fmtSet{$(N'..N]$}};
	\drawArray{15*\boxSize}{0}{12}{2,11}	
	
	\path[draw, bend left , ->] (12 * \boxSize - 0.1, \upperY - 0.1) to ( 9 * \boxSize + 0.1, 0.125);
	\path[draw, bend right, ->] (12 * \boxSize + 0.1, \upperY - 0.1) to (15 * \boxSize - 0.1, 0.125);

,	\node[index label] at ( 0.5*\boxSize, \upperY) {$1$};
	\node[index label] at (23.5*\boxSize, \upperY) {$N$};
	
	\node[index label] at ( -2.5*\boxSize, 0) {$1$};
	\node[index label] at ( 8.5*\boxSize, 0) {$N'$};
	
	\node[index label] at (15.5*\boxSize, 0) {$N'{+}1$};
	\node[index label] at (26.5*\boxSize, 0) {$N$};

\end{tikzpicture} 	\end{center}

	\caption{%
		Sampling~$k$ elements from $[N]$. After splitting $[N]$ and randomly drawing~$X$ from the appropriate distribution (see \cref{sss:uniformSampling}), we can sample~$X$ and $k-X$ items, respectively, from the ---now--- independent subranges in parallel.
	}
	\label{fig:sample_k_from_N}
\end{figure}
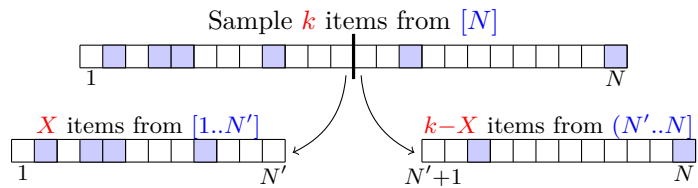
\fi

\indDefKoutOfNSampling%
Sampling~$k$ elements from $[N]$ without replacement is possible in expected time $\Oh{k}$ using a hash table or by sampling skip distances.
In contrast to Bernoulli sampling, it is however necessary to modify the parameters of each new skip distance~\cite{Vit84}.
Sampling without replacement can be parallelized in expected time $\Oh{k/\nproc+\log \nproc}$ using a divide-and-conquer algorithm~\cite{DBLP:journals/toms/SandersLHSD18}.
This algorithm is based on the observation that when splitting a range of size~$N$ into subranges of sizes $N'$ and $N-N'$ respectively (see \Cref{fig:sample_k_from_N}), the number of samples to be taken from the left subrange is distributed hypergeometrically with parameters~$k$, $N'$, and $N$.
This technique can be used to generate $\gGnm$ graphs (\cref{subsec:mod_erdos_renyi}).

By using the binomial distribution instead, one can sample with replacement and this also extends to generating sets of geometric objects in \cref{subsec:mod_geometric_graphs}.
Compared to trivial sampling with replacement, the divide-and-conquer approach has the advantage to allow one processor to generate the objects in some well-defined subspace of the overall sampling space.

\subsubsection{Rejection sampling}\label{sss:rejectionSampling}
\indDefRejectionSampling%
Rejection sampling is a fundamental technique (also known as \emph{acceptance-rejection method}) to draw from a distribution~$\mathcal A$ which lacks an (efficient) direct sampling algorithm.
It requires a second process~$\mathcal B$ that is easy to sample from and that ``overestimates'' $\mathcal A$.
We first sample an element~$x$ from $\mathcal B$, and accept~$x$ with an appropriate probability.
Otherwise, we \emph{reject}~$x$ and repeat the process.
If the acceptance probability is at least a constant, the expected sampling time is of the same order as the sampling time from $\mathcal B$.

For example, suppose we want to randomly draw from $[n]$ where~$i$ should be sampled with probability $p_i$ and where $2 \min_j p_j\geq \max_j p_j$.
Then, we can sample uniformly from $[n]$ and accept the sample with probability $p_i/\max_jp_j\geq 1/2$.
This is expected to succeed with $\Oh{1}$ attempts and succeeds \whp with $\Oh{\log n}$ trials.

\subsubsection{Weighted sampling}\label{sss:weightedSampling}
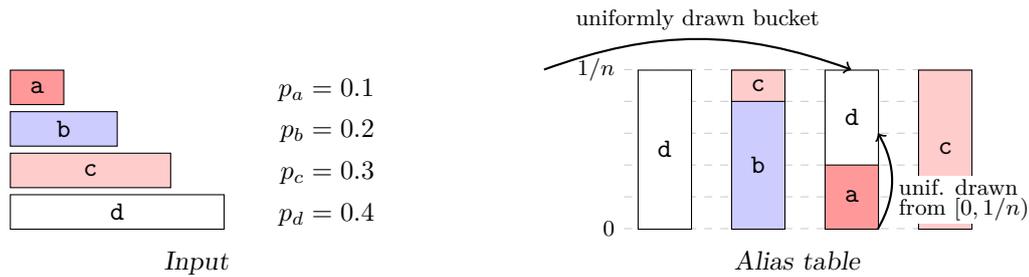
\begin{figure}
   \begin{center}
\begin{tikzpicture}[
   typea/.style={fill=fillColorStrong},
   typeb/.style={fill=fillColor},
   typec/.style={fill=fillColorAlt},
   typed/.style={fill=white}
]
   \node[anchor=north] at (7em, -6.5em) {\textsl{Input}};
   \foreach \v/\x [count=\i] in {0.1/a, 0.2/b, 0.3/c, 0.4/d} {
      \def\py{1.566em -\i * 1.5666em}
      \node[type\x, draw, anchor=north west, minimum width=\v * 20em, minimum height=1.3em, inner sep=0] at (0, \py) {\texttt{\x}};
      \node[anchor=north east, minimum height=1.3em] at (14em, \py) {$p_\x = \v$};
   }

   \def\atX{20em}
   \def\hei{6em}
   \def\wid{2em}
   \def\spa{3.5em}

   \node[anchor=north] at (\atX + 3*\spa - 1em, -6.5em) {\textsl{Alias table}};

   \foreach \y in {0.0, 0.2, 0.4, 0.6, 0.8, 1.0} {
      \path[draw, dashed, black!20] (\atX + 3em, -\hei * \y) to ++(13em, 0);
   }
   \node[anchor=east] at (\atX + 3em, 0) {\footnotesize $1/n$};
   \node[anchor=east] at (\atX + 3em, -\hei) {\footnotesize $0$};

   \foreach \xa/\xb/\lv [count=\i] in {d/d/1, b/c/0.8, a/d/0.4, c/c/1} {
      \def\px{\atX + \spa*\i}
      \node[type\xa, anchor=south west, minimum width=\wid, minimum height=\hei*\lv] at (\px, -\hei) {\texttt{\xa}};
      \ifthenelse{\equal{\lv}{1}}{}{
         \def\py{\hei * \lv - \hei};
         \node[type\xb, anchor=south west, minimum width=\wid, minimum height=\hei - \hei*\lv] at (\px, \py) {\texttt{\xb}};
         \path[draw] (\px, \py) to ++(\wid, 0);
      }
      \node[draw,    anchor=north west, minimum width=\wid, minimum height=\hei] at (\px, 0) {};
   }

   \path[draw, thick, bend left=20, ->] (\atX, 0) to node[above] {\footnotesize uniformly drawn bucket} ++(3*\spa + \wid*0.5, 0);
   \path[draw, thick, bend right, ->] (\atX + 3*\spa + \wid, -\hei) to
   node[right, pos=0.3, align=left, font=\footnotesize, fill=white, inner sep=0.1em, xshift=0.2em] {unif. drawn \\[-0.2em] from $[0, 1/n)$} ++(0, 0.6 * \hei);

\end{tikzpicture}
    \end{center}
   \vspace{-1em}
   \caption{
      Alias table for $n{=}4$ elements $a$, $b$, $c$, and $d$ with weights $p = (1, 2, 3, 4) / 10$.
      The table consists of~$n$ buckets each covering an equal probability mass of $1 / n$.
      Each bucket contains the (partial) probability mass of at most two elements.
      Then, weighted sampling is a two-step process:
      first uniformly select a bucket, then select the element based on the contained partial weights.
   }
   \label{fig:alias_table}
\end{figure}

\indWeightedSampling%
Consider the case where we want to sample from $[n]$ where each element~$i$ appears $w_i$ times.
For small integer weights with $W = \sum_{i=1}^n w_i$, an array~$A[1 .. W]$ in which element~$i$ has a multiplicity of~$w_i$ can be used.
\indPreferentialAttachment%
Sampling an entry from~$A$ uniformly at random yields the required distribution (\cf \cref{subsec:mod_preferential_attachment}).

In the general case, we want to sample from $[n]$ where~$i$ should be sampled with probability $p_i$.
There is a linear time algorithm which computes the data structure depicted in~\Cref{fig:alias_table}.
This so-called \emph{alias table} allows sampling from a discrete distribution in constant time~\cite{walker1977alias,vose1991alias}.
The construction can also be parallelized~\cite{HubSan19}.
The table consists of~$n$ \emph{buckets}, each representing a probability of $1/n$.
The probability of the elements is assigned to the buckets in such a way that each bucket is assigned the probability mass of atmost two elements.
conversely, the mass of some elements may be distributed over multiple buckets.
Sampling then amounts to uniformly sampling a bucket~$i$ and throwing a weighted coin to decide which of the elements assigned to bucket~$i$ is to be returned.

\subsection{Sampling from huge sets}
\label{subsec:hugeSets}

Sometimes we want to sample from a set that is much larger than the output size or the memory of the machine.
For example, we will see several examples where a random graph with~$n$ nodes is defined by probabilities for each entry of an adjacency matrix which has size quadratic in $n$.
In \cref{sss:BernoulliSampling} on Bernoulli sampling we already saw how this works when all the probabilities are the same---we generate skip distances between sampled elements.
Similarly, \cref{sss:uniformSampling} explains how to do it for uniform sampling with or without replacement.
Moreno~\etal~\cite{DBLP:conf/icdm/MorenoPNK14, MorenoPN18} extend this approach to the case when there is only a moderate number of different probabilities -- use one of the above uniform sampling algorithms for each subset of elements with equal probability.

\indRejectionSampling%
We can further generalize this using rejection sampling (\cref{sss:rejectionSampling}).
We partition the data-set into subsets of elements whose probabilities differ only by a constant factor.
In each subset, we perform uniform sampling and use the rejection method to achieve the right sampling probability.
The only prerequisite is that we can compute the exact probability for an element produced by uniform sampling.
\indBernoulliSampling%
\indWeightedSampling%
For example, consider the case where we want to perform weighted subset sampling, i.e., a generalization of Bernoulli sampling where element~$i$ has an individual probability $p_i$. In a subset whose probabilities are between $p/2$ and $p$, we can perform Bernoulli sampling with parameter~$p$ and accept element~$i$ with probability $p_i/p$.

Parallelizing these approaches introduces two levels of parallelism -- coarse-grained parallelization over the subsets with similar probability and fine-grained parallelization within each set using the methods mentioned in \cref{sss:BernoulliSampling,sss:uniformSampling}. For the coarse level, some load balancing is needed since the output sizes for each set heavily depend on the heterogeneous sizes and element sizes in each subset.

\section{Basic Models}\label{sec:gen-models}
\iffalse
	\subsection[Erd\H{o}s-R\`enyi and Gilbert Graphs]{Erdos-Renyi's $\Gnm$ and Gilbert's $\Gnp$ models}
\else
	\subsection[Erd\H{o}s-R\`enyi and Gilbert Graphs]{Erd\H{o}s-Renyi's $\Gnm$ and Gilbert's $\Gnp$ models}
\fi
	\label{subsec:mod_erdos_renyi}
	\label{subsec:mod_gilbert}
\label{sec:erdos_related}
The closely related $\Gnm$ and $\Gnp$ models were the first random graph models considered~\cite{erdds1959random, gilbert1959random}.
They come in different variants that can all be understood as uniform sampling from an $n\times n$ adjacency matrix; see also Sections~\ref{sss:BernoulliSampling} and \ref{sss:uniformSampling}.
\begin{itemize}
	\item If we sample from the whole matrix, we get directed graphs with self-loops (\cf \cref{subsec:special_graphs_directed}).\indGraphDirected
	\item If we exclude the diagonal, we get simple directed graphs.
	\item If we restrict to the upper triangular part of the matrix, we get undirected graphs.
	\item If we exclude suitable blocks, we get bipartite graphs etc.
\end{itemize}
\indBernoulliSampling%
We obtain Gilbert's $\Gnp$ model~\cite{gilbert1959random} using Bernoulli sampling with probability~$p$ for each edge independently.
\indKoutOfNSampling%
If we sample~$m$ edges without replacement, we get the $\Gnm$ model proposed by Erd{\H o}s and R{\'e}nyi~\cite{erdds1959random}.

\indBernoulliSampling%
Batagelj and Brandes~\cite{batagelj2005efficient} present sequential algorithms for these models, and point out several generalizations including a bipartite variant of $\Gnp$.
For $\Gnp$ they propose the Bernoulli sampling approach described in \cref{sss:BernoulliSampling} applied to the upper triangle of the adjacency matrix.
This results in a runtime of $\Oh{n + m}$. For $\Gnm$ they compare two algorithm variants based on hash tables to avoid multi-edges.
\indCommunicationFree\indDistributedMemory%
By now, faster algorithms are available~\cite{DBLP:journals/toms/SandersLHSD18,FUNKE2019200} that are more cache efficient, and also work communication-free in parallel (see also \cref{sss:uniformSampling}).

\indGPU%
Nobari~\etal~\cite{DBLP:conf/edbt/NobariLKB11} proposed a data parallel generator for both the directed and undirected $\Gnp$ model.
\indGraphDirected%
Their generators are designed for graphics processing units (GPUs).
Like the generators of Batagelj and Brandes~\cite{batagelj2005efficient}, their algorithm is based on sampling skip distances but uses precomputations and prefix sums to increase data parallelism.
 	
	\subsection{Preferential Attachment}
	\label{subsec:mod_preferential_attachment}
\indDefPreferentialAttachment%
The preferential attachment model family generates random scale-free networks.
Roughly speaking, when a new vertex is added during the network generation process, it is connected to some existing vertices that are chosen \wrt some of their properties (most often their degrees).
There are multiple ways to generate graphs that follow this framework;
in the following, we describe the \modLongBA (\modBA) model and the \modNodeCopy model.

\subsubsection{Barab\'asi-Albert model}\label{subsubsec:mod_ba}
\iftrue
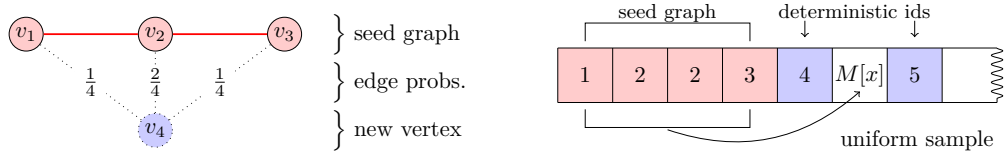
\begin{figure}
	\begin{center}
		\scalebox{0.85}{%
{
	\def\sepx{2}
	\def\sepy{1.5}

	\def\cellw{0.85}
	\def\cellh{0.85}
	\def\arrayy{-1.8*\sepy}

   \begin{tikzpicture}[
      vertex/.style={draw, circle, inner sep=0, minimum width=1.5em, minimum height=1.5em}
   ]
      \node[vertex, fill=fillColorAlt] (v1) at (0, 0) {$v_1$};
      \node[vertex, fill=fillColorAlt] (v2) at (\sepx, 0) {$v_2$};
      \node[vertex, fill=fillColorAlt] (v3) at (2*\sepx, 0) {$v_3$};
      \path[draw, frontColorAlt, thick] (v1) to (v2);
      \path[draw, frontColorAlt, thick] (v2) to (v3);

      \node[vertex, dotted, fill=fillColor] (v4) at (\sepx, -\sepy) {$v_4$};
      \path[draw, dotted] (v4) to node[fill=white] {\large $\frac{1}{4}$} (v1);
      \path[draw, dotted] (v4) to node[fill=white, inner sep=0] {\large $\frac{2}{4}$} (v2);
      \path[draw, dotted] (v4) to node[fill=white] {\large $\frac{1}{4}$} (v3);

      \node[anchor=west] at (2.3*\sepx, 0)      {$\Big\}$ seed graph};
      \node[anchor=west] at (2.3*\sepx, -\sepy/2) {$\Big\}$ edge probs.};
      \node[anchor=west] at (2.3*\sepx, -\sepy) {$\Big\}$ new vertex};
\end{tikzpicture}
\hspace{3em}
\begin{tikzpicture}[
      vertex/.style={draw, circle, inner sep=0, minimum width=1.5em, minimum height=1.5em}
   ]
      \foreach \x/\c in {0/Alt, 1/Alt, 2/Alt, 3/Alt, 4/, 6/} {
           \filldraw[fill=fillColor\c, draw=none] (\x * \cellw, \arrayy) to ++(0, -\cellh) to ++(\cellw, 0) to ++(0, \cellh);
      }

      \foreach \x in {0, ..., 7} {
         \path[draw] (\x * \cellw, \arrayy) to ++(0, -\cellh);
      }
      \path[draw, snake it, segment length=4] (8 * \cellw, \arrayy) to ++ (0, -\cellh);

      \path[draw] (0, \arrayy) to ++(8*\cellw, 0);
      \path[draw] (0, \arrayy-\cellh) to ++(8*\cellw, 0);

      \foreach \l [count=\i] in {1, 2, 2, 3, 4, $M[x]$, 5} {
         \node at (\i * \cellw - 0.5 * \cellw, \arrayy - 0.5 * \cellh) {\l};
      }

      \node[anchor=south] at (2*\cellw, \arrayy + \cellh / 3) {\small seed graph};
      \path[draw] (\cellw / 2, \arrayy + \cellw / 8) to ++(0, \cellh / 3) to ++(3*\cellw, 0) to ++(0, -\cellh/3);

      \node[anchor=south, align=center] (known) at (5.5*\cellw, \arrayy + \cellh / 3)
         {\small deterministic ids\phantom{g}};

      \path[draw, <-] (4.5*\cellw, \arrayy + 0.2*\cellh) to +(0, 0.3*\cellh);
      \path[draw, <-] (6.5*\cellw, \arrayy + 0.2*\cellh) to +(0, 0.3*\cellh);

      \path[draw] (\cellw / 2, \arrayy - \cellh * 1.125) to ++(0, -\cellh / 3) to ++(3*\cellw, 0) to ++(0, \cellh/3);

      \path[draw, ->, bend right] (2* \cellw, \arrayy - \cellh * 1.5) to (5.5 * \cellw, \arrayy - 0.8*\cellw);
      \node[anchor=west] at (5*\cellw, \arrayy - \cellh * 1.7)  {uniform sample};
   \end{tikzpicture}
} %
}
	\end{center}

	\caption{
		\modLongBA generator ($d=1$) by Batagelj and Brandes.
		The seed graph contains nodes $\{v_1, v_2, v_3\}$ and two edges, and we introduce the new node $v_4$.
		Since $v_2$ has two neighbors it is twice as likely to be chosen as the neighbor for~$v_4$.
	}
	\label{fig:ba_bb}
\end{figure}
\fi
Barab\'asi and Albert~\cite{BarabasiAlbert99} define the \modBA model.
It is perhaps most widely used because of its simplicity and intuitive definition:
we start with an arbitrary seed network with nodes $\seq{v_i}{i=1}{n_0}$.
The remaining nodes $\seq{v_i}{i=n_0+1}{n}$ are added one at a time.
They randomly connect to $d$~different neighbors using \emph{preferential attachment}, \ie
the probability to connect~$v_i$ to node $v_j$ is proportional to the degree of~$v_j$ at that time.
The seed graph, $n_0$, $d$, and~$n$ are parameters defining the graph family.

Batagelj and Brandes~\cite{batagelj2005efficient} propose a fast and simple \modBA generator illustrated in \Cref{fig:ba_bb}.
For simplicity of exposition, we use an empty seed graph ($n_0=0$).
A generalization only requires a number of straightforward index transformations.
The algorithm generates one edge at a time and writes it into an edge array $E[1..2dn]$ where positions $2i - 1$ and $2i$ store the node indices of the end points of edge $e_i$ (with $i \ge 1$).
Since each new node has $1 \le d < n_0$ neighbors, we let $E[2i  - 1]= \lceil i/d \rceil$.

The central observation is that one gets the correct probability distribution for the other end point by uniformly sampling from~$E$, \ie $E[2dj+k]$ is set to $E[x]$ where~$x$ is chosen uniformly at random from~$[1..2dj+k)$.
Observe that this formulation allows self-loops and multi-edges.
The former can be prevented by sampling~$x$ from $[1..2dj]$.
To avoid multi-edges one can repeatedly sample until~$d$ different neighbors have been obtained.
Using a hash set of size $\Oh{d}$ and a sufficiently large seed graph, this results only in an expected constant slowdown.

\indIOEfficient%
Meyer and Penschuck~\cite{DBLP:conf/alenex/MeyerP16} propose two I/O-efficient \modBA generators for the external memory model and discuss generalization to various preferential models.
One implements Batagelj's and Brandes' generator using time-forward-processing~\cite{MahZeh-survey} in $\Oh{sort(m)}$.
\indWeightedSampling%
The other one is based on weighted sampling using a tailor-made variant of a Buffer Tree~\cite{DBLP:journals/algorithmica/Arge03} and yields the same I/O-complexity.
The second generator is parallelized by processing subtrees individually.
\indGPU%
Potential bottlenecks near the root are avoided by processing multiple queries to the same tree node in parallel on a GPU.

\iftrue
\begin{figure}
	\begin{center}
		\scalebox{0.93}{%
\begin{tikzpicture}
	\def\boxSize{0.45}

	\def\drawArray#1#2#3{
		\def\xx{#1}\def\yy{#2}\def\n{#3}

		\path[draw] (\xx, \yy) to ++(\n * \boxSize,0);
		\path[draw] (\xx, \yy + \boxSize) to ++(\n * \boxSize,0);

		\foreach \x in {1, 3, ..., \n} {
			\filldraw[fill=fillColor] (\xx + \x * \boxSize - \boxSize, \yy)
			to ++(\boxSize, 0)  to ++(0, \boxSize) to ++(-\boxSize, 0);
		}

		\foreach \x [count=\i] in {0, ..., \n} {
			\path[draw]	(\xx + \x * \boxSize, \yy) to ++ (0, \boxSize);
			\ifthenelse{\equal{\x}{\n}}{}{
				\node at (\xx + \x * \boxSize + \boxSize/2, \yy - \boxSize/2) {\small \i};
			}
		}
	}
	
	\drawArray{0}{-\boxSize}{16}
	\foreach \x [count=\i] in {0, ..., 7} {
		\node at (2*\x * \boxSize + \boxSize / 2, -\boxSize/2) {$v_\i$};
	}

	\foreach \x in {7, 11, 15} {
		\node[opacity=0.5] at (\x * \boxSize + \boxSize / 2, -\boxSize/2) {$v_3$};
	}

	\path[draw, bend right, thick, ->] (15.2*\boxSize, 0.1) to node[above] {\small $h(16) {=} 12$} (11.8*\boxSize, 0.1);
	\path[draw, bend right, thick, ->] (11.2*\boxSize, 0.1) to node[above] {\small $h(12) {=} 8$} (7.8*\boxSize, 0.1);
	\path[draw, bend right=70, thick, ->] ( 7.2*\boxSize, 0.1) to node[above] {\small $h(8) {=} 5$} (4.5*\boxSize, -0.4*\boxSize);

	\filldraw[fill=fillColor] (19*\boxSize, \boxSize * 0.8)
		 to ++(\boxSize, 0)
		 to node[right, align=left, xshift=0.5em] {determinisitc ids of new nodes\\[-0.2em]\ \  $\Rightarrow$ recursion stops} ++(0, \boxSize)
		 to ++(-\boxSize, 0)
		 to ++(0, -\boxSize);

	\path[draw] (19*\boxSize, -1.5*\boxSize)
		to ++(\boxSize, 0)
		to node[right, align=left, xshift=0.5em] {copied from pseudo-random entry\\[-0.2em]\ \  $\Rightarrow$ recursion continues} ++(0, \boxSize)
		to ++(-\boxSize, 0)
		to ++(0, -\boxSize);

\end{tikzpicture} %
}
	\end{center}

	\caption{
		\modLongBA generator by Sanders and Schulz.
		Computation of the edge list's $16$\textsuperscript{th} value: we compute $h(16) = 12$.
		Since the $12$\textsuperscript{th} value was computed by following $h(12) = 8$, we retrace these steps until we hit an odd position.
	}
	\label{fig:ba_ss}
\end{figure}
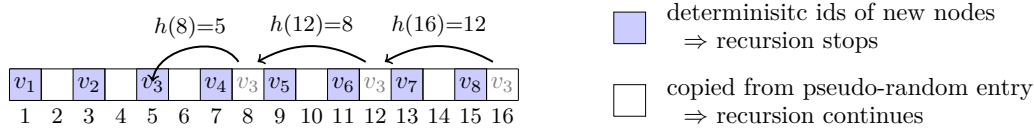
\fi

\indDistributedMemory\indCommunicationFree%
Sanders and Schulz~\cite{DBLP:journals/ipl/Sanders016} propose a communication-free \modBA generator (see \Cref{fig:ba_ss}) based on Batagelj's and Brandes' algorithm.
Recall that  the original algorithm computes the position $M[2i]$ by drawing a random index~$x$ and subsequently copying $M[2i] \gets M[x]$.
The new algorithm, in contrast, obtains a pseudorandom index~$x = h(2i)$ using a random hash function~$h$ satisfying $h(i) < i$ for $i > 1$.
As a result, any processor can at any time of the execution reproduce the value of $x$, and thereby the value of $M[x]$.
If~$x$ is odd the value of $M[x] = \lceil x/2d \rceil$; otherwise, we know that $M[x]$ was obtained by copying from $M[h(x)]$ which we can retrace recursively---again without reading from~$M$.
This process terminates in expected constant time.
As the algorithm never reads from~$M$, all positions can be computed embarrassingly parallel.

\indPowerLaw\indCommunityDetection%
Networks generated with the \modLongBA process have a power law degree distribution~\cite{BarabasiAlbert99},  however, no significant clustering.
Several modifications have been proposed to solve this issue.
Holme and Kim~\cite{PhysRevE.65.026107}, for instance, add another parameter $0 < P_t < 1$.
When adding a new vertex~$u$ with~$d$ links, the first neighbor is added as in \modBA.
For each of the remaining $d{-}1$ neighbors a weighted independent coin is thrown:
with probability $1 - P_t$ preferential attachment is used, otherwise, with probability $P_t$ a triad formation step is carried out.
If edge $\{u, v\}$ was added in the previous preferential attachment step, a neighbor~$w$ of~$v$ is selected randomly, and the edge $\{u, w\}$ is added.
Dorogovtsev and Mendes~\cite{cond-mat/0106144} propose an extreme case;
rather than drawing nodes, their model draws edges and connects a new node to both endpoints, thereby closing a triangle.
Some of the techniques present here are compatible with these modifications.

The parallel generators discussed so far may emit multi-edges.\footnote{%
	In practice, the number of multi-edges is typically small if a sufficiently large seed graph is used. Thus, deleting them does not significantly change the degree distribution.}%
\indDistributedMemory%
The distributed memory generator by Alam~\etal~\cite{AlamKhanMarathe13} (see \cref{subsubsec:mod_copy_model}) produces simple \modBA networks as a special case of the \modNodeCopy model.

\subsubsection{Node Copy Model}\label{subsubsec:mod_copy_model}
\indPreferentialAttachment%
In the \modNodeCopy model~\cite{DBLP:conf/cocoon/KleinbergKRRT99} links to a node are added by picking a random (other) node
in the graph, and copying some links from it.
Similar to the BA model, we start with an arbitrary seed network consisting of $n_0$ nodes and add the remaining $n - n_0$ vertices one at a time.
They randomly connect to~$d$ neighbors using the following process: pick an existing vertex $v_j$ uniformly at random.
Then with probability~$p$, the new node~$v_i$ is connected to $v_j$ (direct edge), and with probability~$1-p$, $v_i$ is connected to a random neighbor of $v_j$ (copy edge).

Alam~\etal~\cite{AlamKhanMarathe13} note that the \modBA model is a special case of the \modNodeCopy model with $p=1/2$.
\indWeightedSampling%
While the opposite is not true, all generators discussed in \cref{subsubsec:mod_ba} can be adapted to the \modNodeCopy model by treating its weighted and unweighted sampling branches individually (see~\cite{DBLP:conf/alenex/MeyerP16} for an in-depth discussion and more generalizations).

A shared-memory parallel algorithm has been proposed by Azadbakht~\etal~\cite{DBLP:conf/sac/AzadbakhtBBA16}. The authors propose an asynchronous parallel method which avoids the use of low-level synchronization mechanisms. Roughly speaking, the process of generating edges can create unresolved dependencies, \ie the corresponding data that needs to be provided by the communicating thread to generate the edge locally has not yet been computed. The authors resolve this problem by having a thread executing two programs. The first part checks if open dependencies are resolved and if so generates the final edge. The second routine tries to generate edges and if there is a dependency the thread stores it to work on it later and continues with the next edge.

\indGPU%
A GPU-based parallel algorithm for the \modNodeCopy model has been presented by~\cite{DBLP:conf/bigdataconf/AlamP17}. In the algorithm, each thread is responsible for a subset of the vertices. In a two phase approach, direct edges and some of the copy edges are created directly.  Due to dependencies, many of the copy edges are put into a waiting queue and created in a second phase, where incomplete edges are resolved and finalized.
The algorithm can generate networks from the model with two billion edges in less than 3 seconds.
\indDistributedMemory%
Alam~\etal~\cite{AlamKhanMarathe13} transfer the algorithm into the distributed memory model and show how dependency chains, which are short in practice, are resolved efficiently in parallel.
Alternatively if multi-edges are acceptable, \cite{DBLP:journals/ipl/Sanders016} can be adapted to multi-GPU scenarios~\cite{alam2019novel} yielding an even more scalable approach.
 
\section{Random Spatial Graphs}
\label{sec:spatial_graphs}
\indDefGraphSpatial%
Entities in real-world networks often have a position in the system that influences their global role in the network as well as their local neighborhood.
The spatial network models presented here account for this phenomenon by associating each vertex with a point placed in some geometric space.
Depending on the model, the probability of connecting two nodes is governed by
(i) their relative orientation (\eg \cref{subsec:mod_geometric_graphs,subsec:mod_delaunay_triangulation,subsec:mod_hyperbolic_graphs})
(ii) their absolute position (\eg \cref{subsec:mod_hyperbolic_graphs}), or
(iii) additional geometric constraints (\eg \cref{subsec:mod_planar,subsec:mod_girg}). 
	\subsection{Random Geometric Graphs}
	\label{subsec:mod_geometric_graphs}
\indGraphSpatial%
\modLongRGGs (\modRGGs)~\cite{gilbert1961random,penrose2003random} are a family of random spatial graphs that project networks onto a $d$-dimensional Euclidean space.
These types of networks have been extensively studied as models for communication networks (\eg ad-hoc wireless networks) and the spreading of diseases~\cite{gilbert1961random}.

\modRGGs are typically constructed by placing~$n$ points uniformly at random in a $d$-dimensional unit-cube $[0,1)^d$.
Subsequently any two points~$x$ and~$y$ with $x \ne y$ are connected by an edge iff their Euclidean distance $d(x,y)$ is within a given threshold radius $r > 0$.
The neighborhood of a node~$x$ thus consists of all points within the $d$-dimensional sphere~$S_x$ with radius~$r$ around $x$.
For nodes near the frontier of the underlying geometry, a significant fraction of~$S_x$ can be outside the unit cube, resulting in lower degrees.
Hence, some variants of \modRGGs use tori rather than cubes to implement periodic boundary conditions (\eg \cref{subsec:mod_girg}).

From percolation theory it is known that for constant $\alpha(d)$ the neighborhood radius $r_c \propto [\ln n /(n \alpha(d))]^{1/d}$ is a sharp threshold on the connectivity of a \modLongRGG with~$n$ points~\cite{penrose2003random}.
For practical purposes one assumes that~$n$ is sufficiently large and~$r$ small.

Waxman~\cite{waxman1988routing} introduces a generalisation of \modRGGs that uses a probabilistic connection function.
In particular, two points $x$, and $y$ are connected with probability $\beta \exp\left[- d(x,y)/(L \alpha)\right]$ where~$L$ is the maximum distance between two points and $\alpha, \beta \in (0,1]$ are parameters controlling the edge density.
To be more specific, a higher value of $\beta$ results in a higher edge density and small values of $\alpha$ increase the density of short edges relative to longer ones.

The trivial bound of $\Oh(n^2)$ work for generating a \modRGG can be improved under the assumption that the points are distributed uniformly at random.
To this end, Holtgrewe~\etal~\cite{DAHoltgrewe} partition the unit-cube into subcubes with side length~$r$.
To find the neighbors of the vertices within a subcube, one only needs to perform distance comparisons between vertices within this cube and the surrounding ones.
This allows the generation of \modRGG in expected time $\symOh(n + m)$.
Holtgrewe~\etal~\cite{DAHoltgrewe,HoltgreweSS10} propose a distributed memory algorithm for the two-dimensional case based on this partitioning.
\indDistributedMemory%
Using distributed sorting and vertex exchanges between processes, they reassign vertices such that edges can be generated locally.
The expected time for the local computation of their generator is $\symOh([n/P] \log(n/P))$, due to sorting.
Perhaps more important for massive-scale systems is that they need to exchange all vertices which yields a communication volume of $\Oh{n/P}$ per process. %

\indDistributedMemory\indCommunicationFree%
A communication-free distributed memory generator for general~$d$ was proposed by Funke~\etal~\cite{DBLP:conf/ipps/FunkeL00SL18}.
Their generator recursively partitions the unit-cube into smaller subcubes by computing a set of binomial random deviates.
\indKoutOfNSampling%
These deviates are sampled consistently across processors similarly to the approach in \cref{sss:uniformSampling}.
In the end, each processor obtains a local subcube and its associated set of vertices.
To generate all adjacent edges for these vertices, each processor redundantly computes the required neighboring subcubes without the need of communication.
The expected number of distance comparisons on each processor then is $\symOh((m+n)/P)$.

Parsonage and Roughan~\cite{DBLP:journals/tnse/ParsonageR17} proposed a fast generator for generalized \modLongRGGs that uses a partitioning approach similar to the one mentioned for plain \modRGG{}s.
Subcubes of the partitioning are used to derive upper bounds on the connection probability for vertices residing in them.
\indRejectionSampling%
Then the methods from \cref{subsec:hugeSets} are applied to each subcube, i.e., candidate edges are identified using Bernoulli sampling which are then filtered using rejection sampling.

One can generate more varied and perhaps more realistic geometric graphs by using real world data to restrict the locations of the points.
For example, one can place points randomly along roads to model wireless car-to-car communication.
To model other ad-hoc wireless networks one could define a probability distribution of points based on known data on population density.
One can also change the definition of edge probabilities in an application specific way.
For example, the \emph{radiation model}~\cite{simini2012universal} has been used to sample trips between locations based on population density.
This can be implemented in a scalable way~\cite{Buchhold0W19} and the result can be viewed as an application specific graph.

	\subsection{Random Hyperbolic Graphs}
	\label{subsec:mod_hyperbolic_graphs}
\indGraphSpatial%
\modLongRHGs~(\modRHG)~\cite{DBLP:journals/corr/abs-1006-5169,DBLP:conf/icalp/GugelmannPP12} are a special case of spatial graphs in which each node has a corresponding point on a two-dimensional\footnote{%
	While empirical studies have been conducted in higher dimensions (\eg~\cite{DBLP:journals/cga/Munzner98}), we are not aware of wide-spread generative models.
} disk~$D$ in hyperbolic space.
Hyperbolic space allows to embed real-world graphs with low distortions~\cite{DBLP:journals/corr/abs-1006-5169};
\indDiameter%
this can be explained intuitively by the fact that the radius of a hyperbolic disk grows logarithmically with its area---just like the diameters of networks often grows logarithmically with their size.
Related to this quirk, the hyperbolic distance $d_H(x, y)$ of points~$x$ and~$y$ does not only depend on their relative position, but also decreases as the pair moves closer to $D$'s center.
Thus, central nodes tend to have above-average many nodes in their vicinity which often renders them network hubs.

In \modRHGs, the point set is scattered randomly onto~$D$ where the radial probability density function increases exponentially towards $D$'s border.
The dispersion parameter $\alpha > 1/2$ controls the density of points near the center which becomes maximal as $\alpha \to 1/2$.
In this case the inner $\Theta(\sqrt n)$ points are expected to form a clique.
For $\alpha = 1$, we obtain a uniform distribution onto hyperbolic space, while for $\alpha > 1$ the density near the center decreases.
As a result, the central clique shrinks to expected constant size and the giant component dissolves~\cite{DBLP:journals/combinatorics/BodeFM15}.

\subsubsection{Threshold model}\label{subsubsec:mod_hyperbolic_graphs_threshold}
The simplest \modRHG variant~\cite{DBLP:conf/icalp/GugelmannPP12} is the \emph{threshold model}.
Here, two points~$u$ and~$v$ are connected iff their hyperbolic distance $d_H(u, v)$ is below the global threshold value~$R$.
\indPowerLaw\indClustering%
With high probability, the resulting graphs have a power law degree distribution with exponent $\gamma = 1 + 2\alpha$, a controllable average degree, a non-vanishing clustering coefficient~\cite{DBLP:conf/icalp/GugelmannPP12}, poly-logarithmic diameter~\cite{DBLP:conf/icalp/FriedrichK15}, and a giant component~\cite{DBLP:journals/combinatorics/BodeFM15} for $\alpha < 1$.

Currently, the most efficient generators for the threshold model share a similar setup originally proposed by~v.~Looz~\etal  \cite{DBLP:conf/hpec/LoozOLM16}.
The algorithm first partitions the hyperbolic disk~$D$ along the radial axis into $\Theta(\log n)$ concentric bands.
Then, it searches the neighbors $\adj{u}$ of each point~$u$ based on a set~$C_u$ of candidate points.
As a crucial optimization, $C_u$ contains only points with a higher radius than~$u$ itself which are computed as follows:
the algorithm overestimates the upper half of the hyperbolic circle~$Q_u$ around~$u$ containing all of $u$'s neighbors.
For each band~$b$ which include a radius identical or large than $u$'s, one computes the smallest and largest angle intersecting $Q_u$ and band~$b$, and accepts all points in between as candidates.
Finally, false-positives in~$C_u$ are filtered out by rejecting all points~$c \in C_u$ with $d_H(u, c) > R$.
Different generators vary in the exact number of bands they use, in the bands' limits as well as how the exact computation of $C_u$ is implemented.

\begin{figure}
	\iffalse \def\nkgenscale{0.7} \else \def\nkgenscale{0.65} \fi
	\begin{center}
		\scalebox{\nkgenscale}{%
\def\rhgCoords{257.7/20.0,7.5/25.0,45.0/41.7,229.6/52.5,179.5/55.0,80.9/55.1,71.3/58.6,273.8/59.1,60.9/63.1,31.8/64.8,246.7/66.9,343.2/67.4,1.4/67.6,184.4/69.0,292.5/69.3,220.5/69.4,259.8/70.5,105.1/73.7,330.4/73.8,257.2/74.7,195.3/75.3,51.2/75.5,134.4/76.0,242.7/76.1,159.1/76.3,156.2/76.7,222.4/77.7,184.7/78.2,234.1/79.1,216.4/79.6,289.9/80.8,187.8/83.1,327.1/83.2,114.9/83.4,32.6/85.0,108.3/85.1,41.0/85.6,298.3/85.6,16.9/86.1,225.5/86.7,197.1/86.8,294.9/87.9,71.6/88.3,308.5/88.6,126.6/88.7,271.7/89.1,106.5/89.2,318.2/89.2,117.2/89.4,59.4/89.6,141.3/90.0,33.6/90.6,295.6/90.8,54.4/90.8,138.3/91.3,339.9/91.6,355.5/91.9,164.3/92.0,297.4/92.1,90.5/92.1,215.1/92.2,325.0/92.3,192.4/92.3,212.5/92.4,14.1/92.6,128.6/92.6,28.7/92.7,110.0/92.9,119.1/92.9,278.6/93.0,14.4/93.3,154.6/94.0,113.4/94.2,229.1/94.5,124.7/95.0,15.5/95.3,316.8/95.4,274.8/95.4,316.1/95.9,150.3/95.9,0.0/95.9,184.8/96.0,215.2/96.1,94.4/96.6,108.3/96.6,9.1/96.6,109.1/96.8,87.1/96.9,200.7/97.0,203.6/97.2,171.0/97.6,105.4/97.7,23.1/97.7,352.4/97.7,122.3/97.8,178.2/98.2,351.7/98.3,158.7/98.9,114.6/98.9,187.1/99.7}%
\def\rhgEdges{1/2,1/3,1/4,1/5,1/6,1/7,1/8,1/9,1/10,1/11,1/12,1/13,1/14,1/15,1/16,1/17,1/18,1/19,1/20,1/21,1/22,1/23,1/24,1/25,1/26,1/27,1/28,1/29,1/30,1/31,1/32,1/33,1/38,1/39,1/40,1/41,1/42,1/44,1/46,1/48,1/53,1/56,1/59,1/61,1/62,1/64,1/70,1/74,1/77,1/78,1/79,1/83,2/3,2/4,2/5,2/6,2/7,2/8,2/9,2/10,2/11,2/12,2/13,2/14,2/15,2/16,2/17,2/18,2/19,2/20,2/22,2/23,2/24,2/31,2/33,2/35,2/37,2/38,2/39,2/42,2/43,2/44,2/48,2/50,2/52,2/54,2/56,2/57,2/62,2/65,2/67,2/71,2/76,2/77,2/79,2/81,2/86,2/93,2/94,2/97,3/1,3/2,3/3,3/4,3/5,3/6,3/7,3/8,3/9,3/10,3/12,3/13,3/18,3/22,3/35,3/37,3/39,3/43,3/50,3/52,3/54,3/65,3/67,3/71,3/76,3/93,4/5,4/8,4/11,4/15,4/16,4/17,4/20,4/24,4/29,4/31,4/38,4/42,4/46,4/70,4/78,5/14,5/16,5/21,5/25,5/26,5/28,5/32,5/41,5/58,5/63,5/82,5/91,5/96,5/100,6/7,6/9,6/10,6/18,6/22,6/43,6/50,6/60,6/84,6/88,7/9,7/10,7/18,7/22,7/43,7/50,7/54,8/11,8/15,8/17,8/20,8/24,8/31,8/46,8/70,8/78,9/10,9/22,9/43,9/50,9/54,10/13,10/22,10/35,10/37,10/39,10/52,10/67,10/93,11/16,11/17,11/20,11/24,11/29,12/13,12/19,12/33,12/56,12/57,12/94,12/97,13/39,13/57,13/65,13/81,13/86,13/94,13/97,14/21,14/28,14/32,14/41,14/63,14/82,14/96,14/100,15/31,15/38,15/42,15/53,15/59,16/24,16/27,16/29,16/30,16/40,16/61,16/64,16/74,16/83,17/20,17/24,17/46,18/34,18/36,18/47,18/68,18/73,18/85,18/87,18/92,19/33,19/56,19/62,20/24,21/28,21/32,21/41,21/63,21/89,21/90,22/37,22/50,22/54,23/45,23/51,23/55,23/66,24/29,25/26,25/58,25/72,25/98,26/58,26/72,26/80,26/98,27/29,27/30,27/40,27/61,27/74,27/83,28/32,28/63,28/82,28/96,28/100,29/40,29/74,30/40,30/61,30/64,30/83,31/38,31/42,31/53,31/59,32/63,32/82,32/100,33/62,34/36,34/49,34/68,34/69,34/73,34/87,34/99,35/37,35/52,35/67,36/47,36/68,36/73,36/85,36/87,36/92,38/42,38/53,38/59,39/65,39/71,39/76,40/74,41/63,41/89,42/53,42/59,45/66,45/75,45/95,46/78,47/68,47/85,47/87,47/92,48/77,48/79,49/69,49/73,49/99,50/54,51/55,52/67,53/59,57/81,57/94,57/97,60/84,60/88,61/64,61/83,64/83,65/71,65/76,66/75,68/73,68/85,68/87,69/95,70/78,71/76,72/80,73/99,75/95,77/79,81/2,81/13,81/57,81/81,82/100,85/87,85/92,87/92,89/90,94/97}%
\def\rhgDTs{91.0/55.0,103.5/58.4,113.6/61.9,122.2/65.3,129.4/68.8,135.6/72.2,141.0/75.7,145.7/79.1,149.7/82.6,153.3/86.0,156.4/89.5,159.2/92.9,161.6/96.4,163.8/99.8,165.7/103.3,167.3/106.7,168.8/110.1,170.1/113.6,171.3/117.0,172.3/120.5,173.2/123.9,174.0/127.4,174.8/130.8,175.5/134.3,176.1/137.7,176.7/141.2,177.2/144.6,177.7/148.1,178.3/151.5,179.5/155.0,179.5/155.0,180.6/151.5,181.2/148.1,181.7/144.6,182.3/141.2,182.8/137.7,183.5/134.3,184.1/130.8,184.9/127.4,185.7/123.9,186.6/120.5,187.7/117.0,188.8/113.6,190.1/110.1,191.6/106.7,193.3/103.3,195.2/99.8,197.3/96.4,199.7/92.9,202.5/89.5,205.6/86.0,209.2/82.6,213.3/79.1,217.9/75.7,223.3/72.2,229.5/68.8,236.8/65.3,245.3/61.9,255.4/58.4,267.9/55.0}%
\xdef\distPath{0}%
\foreach \p/\r [count=\i] in \rhgDTs {%
	\ifthenelse{\equal{\i}{1}}{%
		\xdef\distPath{(1pt*\p, 1.5pt*\r)}%
	}{\ifthenelse{\lengthtest{\r pt < 105pt}}{%
			\xdef\distPath{\distPath -- (1pt*\p, 1.5pt*\r)}%
		}{}}}%
\newcommand{\hypCS}[4]{
	\node[draw, minimum height=#4, minimum width=#3, thick] (#1) at (#2) {};
	
	\node[clabel, anchor=east] at (#1.north west) {$R$};
	\node[clabel, anchor=east] at (#1.south west) {$0$};
	
	\node[clabel, anchor=north] at (#1.south west) {$0$};
	\node[clabel, anchor=north, yshift=-0.2em] at (#1.south) {angle~$\vartheta$};
	\node[clabel, anchor=north] at (#1.south east) {$2\pi$};
}
\begin{tikzpicture}[
	every node/.style={font=\Large},
	clabel/.style={},
	rgm back/.style={draw, circle, goetheblau, thick, minimum height=200pt,fill=sandgrau!50},
	point/.style={cross out, draw},
	dist circle/.style={circle, emorot, dashed, draw, minimum width=100pt},
]
	\hypCS{r}{180pt,75pt}{360pt}{150pt}
	\foreach \y in {50, 90, 120, 140} {
		\path[draw, black, dotted] (0, 1pt*\y) to (360pt, 1pt*\y);
	}

	\node[anchor=east] (band-label) at (-55pt, 75pt) {bands};
	\foreach \y in {25, 70, 105, 130, 145} {
		\path[draw, -o] (band-label) to (10pt, \y pt);
	}

	\foreach \y/\h/\r/\left/\right in {50/40/82.5/91.0/267.9, 90/30/90/109/250, 120/20/120/146/212, 140/10/140/159/199} {
		\node[inner sep=0, anchor=south west, frontColorAlt, fill=fillColorAlt, draw, minimum height=1pt*\h, minimum width=1pt*(\right-\left)] at (1pt*\left, 1pt*\y) {};
	}

	\draw[frontColor, fill=fillColor] \distPath;
	\path[dashed, thick, draw] (150pt, 155pt) to (210pt, 155pt);

	\foreach \y/\h/\r/\left/\right [count=\i] in {50/40/82.5/91.0/267.9, 90/30/90/109/250, 120/20/120/146/212, 140/10/140/159/199} {
		\node[point, frontColor, thick, inner sep=3pt, rotate=45] (expt-left-\i) at (1pt*\left, 1pt*\r) {};
		\node[point, frontColor, thick, inner sep=3pt, rotate=45] (expt-right-\i)	at (1pt*\right, 1pt*\r) {};
	}

	\foreach \p/\r [count=\i] in \rhgCoords {
		\ifthenelse{\equal{\i}{5}}{
			\node[point, inner sep=3pt] (p\i) at (1pt*\p, 1.5pt*\r) {};
			\node[anchor=north west] at (1pt*\p, 1.5pt*\r) {$q$};
		}{
			\ifdim\p pt<260pt
				\node[point, inner sep=1.5pt] (p\i) at (1pt*\p, 1.5pt*\r) {};
			\fi
		}
	}

	\node[align=left, anchor=west] (cand-label) at (370pt, 140pt) {overstimated search area\\[-0.2em]containing the candidates};
	\node[align=left, anchor=west] (true-label) at (370pt, 90pt) {upper half of hyperbolic\\[-0.2em] circle around $q$ containg\\[-0.2em] the true neighbors};
	
	\node[align=left, anchor=west] (extrem-label) at (370pt, 30pt) {largest angle at which\\[-0.2em] neighbors may be \\[-0.2em] positioned in this band};
	
	\path[draw, -o] (cand-label.west) to (240pt, 110pt);
	\path[draw, -o] (true-label.west) to (220pt, 100pt);
	\path[draw, -o] (extrem-label.west) to (266pt, 83.5pt);

	\foreach \i/\j in \rhgEdges {
	\ifthenelse{\equal{\i}{5}}{
	  \path[draw, frontColorStrong, thick] (p\i.center) to (p\j.center);
	}{
	  \ifthenelse{\equal{\j}{5}}{
	    \path[draw, frontColorStrong, opacity=0.3, dashed] (p\i.center) to (p\j.center);
	  }{}
	}
	}

	\node[clabel, anchor=south, rotate=90, yshift=0.2em, fill=white, opacity=0.9, inner sep=0.2em] at (0, 75pt) {\phantom{radius~$r$}};
	\node[clabel, anchor=south, rotate=90, yshift=0.2em] at (0, 75pt) {radius~$r$};
\end{tikzpicture}
}
	\end{center}
	\vspace{-1em}
	\caption[NkBand's partitioning]{
		\algNkBand partitions the Hyperbolic disk (here in polar coordinates) into radial bands.
		For each point~$q$, it searches neighbors in the $q$'s own band and all above.
		To this end, we overestimate the upper half of the hyperbolic circle around~$q$ by computing the left and right most angles intersections between the hyperbolic circle and the bands.}
	\label{fig:nkgen}
\end{figure}
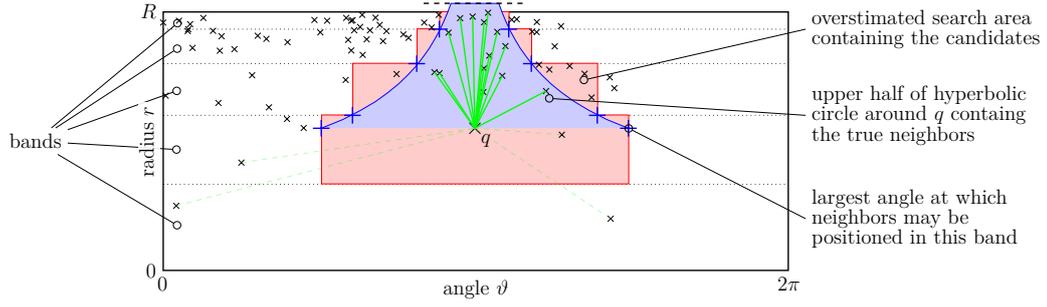

\algNkBand by v.~Looz~\etal~\cite{DBLP:conf/hpec/LoozOLM16}  initially draws the complete point set.
As illustrated in~\Cref{fig:nkgen}, each band is effectively an array storing all points contained sorted by their angles.
To compute $C_u$, a binary search for the left- and right-most points is carried out in each relevant band.
The authors demonstrate an empirical runtime of $\Oh{n\log n + m}$.
The parallel implementation is available as part of \implNetworKit.

\indStreaming%
Penschuck~\cite{DBLP:conf/sea/Penschuck17} proposes the streaming generator \algHyperGen using a small memory footprint with a cache-aware implementation in mind.
The generator overlaps the sampling of points with the neighborhood search in a sweep-line algorithm resulting in a memory footprint of $\Oh{[n^{1- \alpha} \bar d^\alpha + \log n] \log n}$ \whp or a time complexity of $\Oh{n\log\log n + m}$ \whp.

\indDistributedMemory\indCommunicationFree%
Funke~\etal~\cite{DBLP:conf/ipps/FunkeL00SL18} introduce the communication-free generator \algRHG as part of \implKaGen.
It further sub-divides bands into cells which can be independently recomputed by all processors without communication (see \cref{subsec:prelim-basic-sampling-tech}).
Then, for each point~$u$ the set of candidates~$C_u$ consists of all points in cells intersecting the hyperbolic circle~$Q_u$ around~$u$.
The generator requires expected time $\Ohsmall{(n+m)/P + P \log n + n (P\bar d/n)^\alpha  + n^{\frac{1}{2\alpha}}}$, where~$P$ is the number of processors.

Combining the streaming technique of~\cite{DBLP:conf/sea/Penschuck17} with the communication-free sampling of~\cite{DBLP:conf/ipps/FunkeL00SL18}, Funke~\etal~\cite{FUNKE2019200} propose \algsRHG capable of generating a graph with $2^{39}$ nodes and $2^{42}$ edges on $p = 2^{15}$ cores in 1~min.

\subsubsection{Binomial model}\label{subsubsec:mod_hyperbolic_graphs_general}
Similarly to \modRGG, there exists a generalization of \modRHG that replaces the sharp distance threshold of connected nodes by a distance dependent connection probability.
The so-called \emph{binomial} \modRHG features an additional\footnote{%
	The original proposal~\cite{DBLP:journals/corr/abs-1006-5169} by Krioukov~\etal already included this parameter, as well as an additional parameter $\zeta$ which does not add a degree-of-freedom and is omitted here without loss of generality.}
temperature parameter~$T \ge 0$ controlling the sharpness of the decision threshold.
This gives rise to various parameter regimes~\cite{DBLP:journals/cphysics/AldecoaOK15}.
For $T=0$, we obtain the threshold model.
For $T > 0$ edges between two points~$u$ and~$v$ are created with a probability that decreases exponentially with $d_H(u, v)$.
In the extreme of $T \rightarrow \infty$, the model degenerates into $\Gnp$.

\indQuadTree%
The first generator with sub-quadratic work is \algNkQuad by v. Looz~\etal~\cite{DBLP:conf/isaac/LoozMP15}.
It stores the points, which are projected on to a Poincar\'e disk, in a polar quad-tree.
Then, the query of point~$u$ samples leaves of the quad-tree by bounding the connection probability to connect to a point in such a leaf from above.
All points within a leaf are treated as candidates and randomly selected.
The generator has a worst-case runtime of $\Oh{ (n^{3/2} + m) \log n }$.
The parallel implementation is available as part of \implNetworKit.

Later, v. Looz~\etal~\cite{looz18thesis} improve these results using a band-based partition as in \algNkBand.
For long-ranged edges the algorithm exploits that the probability of an edge decreases monotonously as the distance between its endpoints increases.
Hence, it uses a combination of geometric jumps and rejection sampling as discussed in \cref{subsec:hugeSets}.
\indRejectionSampling%
The resulting generator has an expected runtime of $\Oh{n \log^2 n + m}$.

Bl\"asius~\etal~\cite{DBLP:conf/esa/Blasius0K0PW19} propose \algHyperGIRGs, a generator with expected linear work for $T < 1$ based on the \modGIRG model.
While \algNkQuad and \algHyperGIRGs are conceptually very similar, \algHyperGIRGs operates in the native geometry and navigates the quad-tree more efficiently (see \cref{subsec:mod_girg}).
 	
	\subsection{Geometric Inhomogenous Random Graphs}
	\label{subsec:mod_girg}
\indGraphSpatial%
Brinkmann~\etal~\cite{DBLP:journals/tcs/BringmannKL19} propose the \modLongGIRG (\modGIRG) model.
It identifies each node with a point on a $d$-dimensional torus and---similarly to the \modLongChungLuRef model---assigns each node a non-negative weight.
The probability of an edge to be added between two nodes is then a function of the points' distance and their weights.
The authors show that general \modRHGs are asymptotically contained in the $(d{=}1)$-dimensional \modGIRG model if hyperbolic radii are projected to \modGIRG weights, and angles are mapped onto the $1$-dimensional torus.

\indQuadTree%
There exists a sampling algorithm~\cite{DBLP:journals/tcs/BringmannKL19} with expected linear work.
It decomposes the geometry similarly to a $d$-dimensional quad-tree and associates intervals of the node weight with the tree layers.
To this end, nodes with many potential neighbors (i.e., a high weight) are placed near the tree's root which makes them candidates to a large subset of the underlying torus.
The generator arranges the tree's leaves in memory using a space filling curve which allows to efficiently iterate over all points contained in arbitrary subtrees.

Bl\"asius~\etal~\cite{DBLP:conf/esa/Blasius0K0PW19} engineer the two generators \algHyperGIRGs and \algGIRGs.
While the latter samples multi-dimensional \modGIRG graphs, \algHyperGIRGs is slightly modified to sample from the exact \modRHG probability distribution and can efficiently generate \modRHG instances with $T < 1$.
At time of writing, the parallel implementation of \algHyperGIRGs is the fastest sequential generator for the threshold and binomial model.
 
	\subsection{Random Planar Graphs}
	\label{subsec:mod_planar}
\indDefGraphPlanar%
Planar graphs can be drawn in the plane such that no edges cross.
As they are an intensively studied family of graphs, generators for random planar graphs are very interesting.
Perhaps the most natural model asks for a uniform sample from all planar graphs with a given number of nodes~$n$.
Such graphs can be sampled in expected quadratic time.
If the network size must be realized only up to a constant factor, expected linear time sampling is possible~\cite{fusy2009uniform}.
Although this algorithm has been implemented to generate graphs with $10^5$ nodes, it is not very well suited for scaling to much larger graphs since heavy precomputations are needed.

Several models of planar graphs that are easier to generate have been considered~\cite{meinert2011experimental}.
The most scalable of these models consider Delaunay triangulations of random point sets (\cref{subsec:mod_delaunay_triangulation}). 	
	\subsubsection{Random Delaunay Triangulations}
	\label{subsec:mod_delaunay_triangulation}
\indDelaunayTriangulation%
\indGraphPlanar\indGraphSpatial%
The \textsl{Delaunay triangulation} (DT) of a point set~$P$ partitions $P$'s convex hull into triangles such that no circumcircle of a triangle contains a forth point in $P$.
This so-called \emph{Delaunay condition} maximises the minimal angle in each triangle and typically avoids near-degenerate triangles with very sharp angles.
There are several generators that generate such a triangulation of a random point set in the unit square (or cube).
This is an appealing family of instances since two-dimensional DTs can be generated efficiently.
Additionally, the resulting graphs resemble meshes used in numerical computations.
Indeed, DT is an important ingredient in generating meshes for numerical simulation (\eg~\cite{shewchuk2002delaunay}).
For example, for graph partitioning, these graphs are interesting instances since fluctuations in point densities mean that non-trivial partitions are sought that steer through thin areas of the graph.

The widely used sequential two-dimensional generator by Holtgrewe~\cite{DAHoltgrewe,HoltgreweSS10}  chooses each point independently and uniformly at random.
Since the node numbering implies no locality at all, such instances are unexpectedly difficult for parallel algorithms.
To expose more locality, the parallel generator by Funke~\etal~\cite{DBLP:conf/ipps/FunkeL00SL18} uses the same local point generation strategy discussed in \modLongRGGRef.
\indDistributedMemory\indCommunicationFree%
Moreover, to enable a simple, communication-free, and highly scalable parallelization, this generator uses periodic boundary conditions, \ie distances are computed as the smallest Euclidean distance to any copy of the point set in $x$-, $y$-, (or $z$-)direction.
A periodic boundary condition allows them to avoid long Delaunay edges, thereby reducing the number of recomputations.
Note that this model adaptation is also practically relevant since periodic boundary conditions appear in many scientific simulations.
The generator is freely available as part of \implKaGen, and supports a three-dimensional variant of the Delaunay generator. 	
	\subsubsection{Planar Graphs for Infrastructures}
	\label{subsec:mod_planar_infra}
\indGraphPlanar%
Planar graph generation has always attracted attention of engineers because graphs underlying many types of civil infrastructure are either completely or almost planar. Examples include road networks, power grids, water distribution systems, and natural gas pipelines. Attempts to use random planar graph generators such as Plantri~\cite{brinkmann2007fast}, Fullgen~\cite{brinkmann2011program}, and Markov Chain based~\cite{denise1996random} have not ended up with a desired realism even if the generated graphs have been refined to better fit desired properties. Because there is a shortage of high-resolution real data for these networks (except the road networks such as the OpenStreetMap) due to various reasons, such as cost of information collection and confidentiality, generation of domain-specific planar graphs is of particular interest.

Because most of these generators are developed for practical purposes, they rely on domain-specific properties.
\indClustering\indGraphSpatial%
For example, road network generators may consider a realistic population density modeled using clustering effects fused with geometric graph edge generation rules and subsequent edge rewiring~\cite{hackl2017generation} or a hierarchical city-town-village structure of nodes with domain-specific proximity based rules for edges~\cite{galin2011authoring}.
Power grids are typically spatially-defined and nearly-planar graphs.
Despite the fact that many models are claimed to capture the structure of power grids without specific planarity requirement, they either require a planarization postprocessing or model adjustments to generate realistic graphs that can represent a power grid.
For example, in~\cite{wang2008generating}, the nodes and edges are generated using various random distribution functions within small fixed areas, and proximity constraints, respectively.
In~\cite{aksoy2018generative}, the \modLongChungLu model is initializing the backbone of the graph, and random star-like structures are added to it. Both combinations seem to generate nearly planar graphs. In another domain, a water distribution system is generated by randomly concatenating small grid graphs taking into account domain specific constraints~\cite{sitzenfrei2013automatic}.

\section{Random Graphs with Prescribed Degree Sequences}
\indBeginDegreeSequence%
\label{sec:graph-randomization}
Sampling graphs with a prescribed degree-sequence is a classical problem in theoretical computer science with many practical applications (\eg as a building block in \modLFRRef benchmark).
\indDegreeSequence%
The task is, given a sequence of degrees $\degseq = \seq{d_i}{i=1}{n}$, to return a uniform sample from the ensemble of all graphs where each node $v_i$ has degree $d_i$.
The problem is intimately related with the challenge of uniformly perturbing the edges of an existing graph which is frequently used in network analysis for hypothesis testing (\eg~\cite{Milo824, gotelli1996null}).

The \modLongCMRef yields a linear time algorithm that may produce graphs with self-loops or multi-edges.
Applications, however, often require \emph{simple} graphs without those structures.
Sampling from such an ensemble is more involved and not all degree sequences can yield a simple graph;
we call a degree sequence~$\degseq$ \emph{graphical} iff it does.
\indDefDegreeSequenceGraphical%
A fast approximate solution can be obtained using the model by \modLongChungLuRef which realizes $\degseq$ only in expectation.
Alternatively, one can generate (potentially) non-simple graphs using the \modLongCM and subsequently deal with forbidden edges (see \cref{subsec:rand_configuration_model}).
In contrast, the frequently used \modLongFDSM (\modFDSM) directly yields simple graphs in a two-step approach.
It first generates a highly biased graph in linear time (e.g., using the \algHavelHakimi~\cite{Havel1955, doi:10.1137/0110037} and engineered by~\cite{DBLP:journals/jea/HamannMPTW18}), and then perturbs the instance using a sufficiently long sequence of small local updates (\eg \modLongESRef and \modLongCBRef). 
	\subsection{Chung-Lu}
	\label{subsec:mod_chung_lu}
\indDegreeSequence%
Chung and Lu (\modChungLu)~\cite{chung2002connected} describe random graphs designed to match a prescribed degree sequence $\degseq$ in expectation.
\modChungLu graphs are parameterized by an $n$-dimensional non-negative vector~$w = \seq{w_i}{i=1}{n}$.
In order to be realizable, $w$'s largest value has to be restricted to $\max_i w_i^2 \le W$ where $W \Def \sum_i w_i$.
Then, each node $v_i$ is assigned the weight $w_i$ and two nodes~$v_i$ and $v_j$ are connected with probability $p_{ij} := w_i w_j /W$.
While one typically chooses $w = \degseq$, the vector may also contain real-valued entries.

Miller and Hagberg~\cite{DBLP:conf/waw/MillerH11} give the first efficient generator capable of producing simple \modChungLu instances.
Their sequential algorithm requires a non-increasing weight sequence.\footnote{%
	This restriction can be lifted by appropriately sorting and relabeling nodes.}
As a result, the probability~$p_{ij}$ of an edge between a fixed node~$v_i$ and a partner $v_j$ only decreases as~$j$ increases, i.e. we have $p_{ij} \ge p_{ik}$ for all $j < k$.
Thus, after considering the edge $(v_i, v_j)$ we can obtain the next candidate $(v_i, v_k)$ by sampling a geometric skip distance with $p_{ij}$ and correct the potentially overestimated probability by accepting the candidate only with probability $p_{ik} / p_{ij}$ (\cf \cref{subsec:hugeSets}).
The resulting generator has an expected runtime of $\Oh{n+m}$.
The approach can be parallelized using the approach used in KaGen~\cite{DBLP:conf/ipps/FunkeL00SL18}---the adjacency matrix is partitioned into slices which can be generated independently.
Appropriate load balancing has to take into account that different parts of the matrix incur a highly different amount of work.

Alam~\etal~\cite{DBLP:conf/sc/AlamKVM16} also require a sorted weight sequence which they collapse into $N_w$ groups each containing nodes with identical weight.
For each group only $\Oh{1}$ words are stored.
Then, they conceptually decompose the adjacency matrix into $\Theta(N_w^2)$ blocks where each block is sampled as a bipartite \Gnp{} graph.
The authors show a runtime of $\Oh{N_w^2 + m}$ and demonstrate an speed-up over~\cite{DBLP:conf/waw/MillerH11} for practical weight sequences.
They also give a parallel variant requiring time $\Oh{(m + N_w^2)/ P + P + N_w}$ where~$P$ is the number of processors.

Moreno~\etal~\cite{MorenoPN18} unify both approaches and generalize them into an algorithmic framework suited for static graph models where each edge $\{u,v\}$ independently exists with an (implicitly defined) probability $p_{uv}$ and the number $\left|\{p_{uv}\, |\, u,v \in V \}\right|$ of unique probabilities is small.
Then, each group of edges with identical probabilities is treated separately.
The authors describe two sampling strategies both yielding an expected sequential time of $\Oh{n + m + N_w^2}$ where $N_w$ is the number of unique node weights.\footnote{%
	Observe that in the common case where~$w$ contains integers, the $N_w^2$ terms in the previous results are asymptotically dominated by the expected number of edges.
	Let~$G$ be a graph containing $N_w$ different degrees and let $V' \subset V$ be an arbitrary set of nodes, s.t. all nodes in $V'$ have a different degree.
	The nodes in $V'$ are incident to at least $\sum_{v \in V'} \degree(v) / 2 \le \sum_{i = 0}^{N_w - 1}i / 2 = \Theta(N_w^2)$ edges.
	Hence, we expect $N_w^2 = \Oh{m}$ which also holds \whp for $N_w = \Omega(\log n)$ by bounding the realized degrees from below using Chernoff's inequality.
}

The dependence on $N^2_w$ can be removed by applying the techniques from \cref{subsec:hugeSets}.
Concretely, we can subdivide the set of adjacency matrix entries into a logarithmic number of groups such that the entries within a group have probabilities that differ by at most a constant factor.
Sampling within a group can then be done using a combination of Bernoulli sampling and rejection sampling.
\indRejectionSampling\indBernoulliSampling%
Although this may demand a small constant factor more calculations then the other approaches, it can be implemented very efficiently because the involved computations (generating skip distances and making acceptance decisions) are very simple and have high data parallelism.
Thus, they can use parallelism, vector instructions, GPUs, and existing highly tuned library codes for these tasks. 
	\subsection{Configuration model}
	\label{subsec:rand_configuration_model}
The \modLongCM{} (\modCM) was initially conceived for the theoretical analysis of random graphs~\cite{DBLP:journals/jct/BenderC78,newmann10,bollobas1985random}.
Due to its simplicity, it is now also used to sample from prescribed degree sequences~\cite{DBLP:journals/rsa/MolloyR95a}, and among others motivated the notion of \emph{modularity}.

\indDegreeSequence%
In the following, we consider the \modLongCM for undirected graphs, and introduce a directed variant in \cref{subsec:special_graphs_directed}.
Given a degree sequence $\degseq = \seq{d_i}{i=1}{n}$ with $\sum_i d_i = 2m$, \modCM generates a graph $G(V, E)$ with $\card{V} = n$ and $\card E = m$.
We first create an urn~$U$ which contains exactly $d_i$ balls labeled~$v_i$ for each node $v_i \in V$.
Then, we draw and remove two balls from~$U$ uniformly at random, and add the edge $\{u, v\}$ to~$E$ where~$u$ and~$v$ denote the labels of the two balls respectively.
The process terminates when~$U$ is empty.

Clearly, \modCM allows for self-loops and multi-edges\footnote{%
	Due to self-loops and multi-edges, not all graphs have the same probability to be generated~\cite[p.436]{newmann10}.
	Hence, \modCM is not guaranteed to produce uniform samples.
}.
While their expected number is small for any graph~$G$ with $\maxdeg(G) \ll n$, multiple strategies to obtain simple graphs have been considered:

\begin{itemize}
	\item The \modLongECM deletes all self-loops and multi-edges without replacement, and thereby changes the degree sequence non-uniformly.
	This can result in significant structural changes~\cite{DBLP:journals/snam/SchlauchHZ15}.
	
	\item \emph{Rejection sampling} repeatedly samples using the \modCM, and accepts the first simple instance obtained.
	\indRejectionSampling%
	This method, however, only yields expected polynomial runtimes if the maximum degree $\maxdeg(G) = o(\sqrt{\log n})$ is small.
	
	\item Recent theoretical results by Arman~\etal~\cite{DBLP:conf/focs/ArmanGW19} (building on~\cite{DBLP:journals/siamcomp/GaoW17, DBLP:journals/jal/McKayW90}) improve upon rejection sampling by transforming a multi-graph sampled with \modCM{} into simple graphs---one defect at a time.
	The technique is very similar to the one sketched for \algIncRegRef, but uses more switches and treats high degree nodes separately.
	
	\item \emph{Random rewiring steps} are a common technique to remove forbidden structures (\eg~\cite{Lancichinetti2009}) in a Markov-style Las Vegas algorithm.
	Hamann~\etal~\cite{DBLP:journals/jea/HamannMPTW18} use such pseudo-random edge swaps (\cf \cref{subsec:rand_edge_switching}) at scale to heuristically remove the unwanted edges while preserving the original degrees.
\end{itemize}

\indRandomPermutation%
Efficient implementations of \modCM typically exploit that by removing random balls from the urn until it is empty, the model effectively establishes a random permutation of all balls initially contained.
In fact, our description of \modCM is very similar to the \algFisherYatesShuffleRef.
Hence, the following reformulation is equivalent, and leads to scalable generators:
store the contents of urn~$U$ as a sequence~$S$, shuffle~$S$ to obtain a permutation uniformly at random, and then interpret~$S$ as a sequence of~$m$ node-pairs encoding the edge set~$E$.
 	
	\subsection{Edge switching}
	\label{subsec:rand_edge_switching}

\iftrue
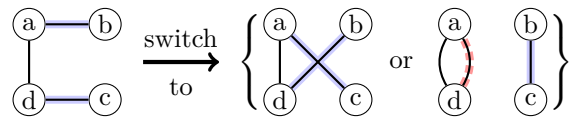
\begin{figure}
	\begin{center}
\begin{tikzpicture}[
	vertex/.style={draw, circle, inner sep=0, minimum width=1.2em, minimum height=1.2em},
	edge/.style={draw, thick},
	highlighted edge/.style={draw, line width=3, fillColor}
]

\foreach \x [count=\i] in {1, 4.3, 6.6} {
	\node[vertex] (a\i) at (\x  , 0.5) {a};
	\node[vertex] (b\i) at (\x+1, 0.5) {b};
	\node[vertex] (c\i) at (\x+1,-0.5) {c};
	\node[vertex] (d\i) at (\x  ,-0.5) {d};
}

\path[highlighted edge] (a1) to (b1);
\path[highlighted edge] (c1) to (d1);
\path[edge] (c1) to (d1) to (a1) to (b1);

\path[highlighted edge] (a2) to (c2);
\path[highlighted edge] (b2) to (d2);
\path[edge] (a2) to (c2);
\path[edge] (b2) to (d2);
\path[edge] (d2) to (a2);

\path[highlighted edge] (a2) to (c2);
\path[highlighted edge] (b2) to (d2);
\path[edge] (a2) to (c2);
\path[edge] (b2) to (d2);

\path[highlighted edge] (b3) to (c3);
\path[edge] (b3) to (c3);

\path[edge, bend left]  (d3) to (a3);
\path[highlighted edge, bend right, fillColorStrong, dotted]  (d3) to (a3);
\path[edge, bend right] (d3) to (a3);

\node at (3.9, 0) {\Large $\Bigg\{$};
\node at (5.9, 0) {or};
\node at (8.0, 0) {\Large $\Bigg\}$};

\path[draw, ultra thick, ->] (2.5, 0) to node[align=center] {switch\\[0.5em]to} (3.5, 0);
\end{tikzpicture}
 	\end{center}

	\caption{%
		Edge Switching in an undirected graph.
		Two edges (here $\{a,b\}$ and $\{c,d\}$) and one of two possible new topologies are drawn uniformly at random.
		The swap is rejected, if the new induced subgraph is not simple (right example).
	}
	\label{fig:edge_switching}
\end{figure}
\fi

The \modLongES (also known as re-wiring, swap, or trade) model~\cite{rao1996markov,Mihail2003} applies a succession of~$k$ small perturbations, so-called edge switches, to a network.
In case of an undirected input (see \Cref{fig:edge_switching}), two random edges $\{u, v\}$ and $\{x, y\}$ are replaced either by $\{u, y\}$ and $\{x, v\}$, or---equiprobably---by $\{u, x\}$ and $\{v, y\}$.
Clearly, these changes preserve the degrees of all nodes involved.
If a swap violates application-specific constraints (\eg by introducing a self-loop into a simple graph), it is rejected without replacement.

This random process is often modeled as a Markov-chain:
\indDegreeSequence%
the state space corresponds to the ensemble of all legal graphs with the same degree sequence.
We connect two states iff their graphs can be transformed into each other using a single swap.
For directed graphs and simple undirected graphs, the Markov chains have symmetric transition probabilities, are irreducible and aperiodic;
hence, a sufficiently long sampling process converges to a uniform sample~\cite{CarstensPhd}.\footnote{%
	Carstens~\cite[section~2.2]{CarstensPhd} discusses adaptations for additional graph classes; see also~\cite{carstens1803.02624}.
}
While rigorous upper bounds on mixing times are only known for special graph classes and are impractically high (\eg~\cite{DBLP:journals/corr/abs-1903-06600}), in practice a constant number of swaps per edge often yields sufficiently uniform results~\cite{Mihail2003, DBLP:journals/jea/HamannMPTW18, CarstensPhd}.

\indGraphDirected%
In the following, we only report on the randomization of simple undirected graphs since the directed variant implies fewer constraints and is significantly easier.
Swapping of simple undirected graphs requires a data structure that supports the following steps efficiently: (i) gather two random edges uniformly at random, (ii) test whether their replacements already exists (to prevent multi-edges), (iii) update the selected edges.
Step~(ii) implies that the neighborhoods of up to four nodes have to be considered.

Viger and Latapy~\cite{DBLP:journals/corr/abs-cs-0502085} implement a graph data structure similar to an adjacency list where each neighborhood $N(u)$ is stored as a hashset;
small $N(u)$ are kept in arrays as an optimization.
Executing $k$~swaps then requires expected $\Oh{k}$ work, but causes unstructured memory access resulting in a significant slow-down for large graphs.

\indIOEfficient%
Hamann~\etal\cite{DBLP:journals/jea/HamannMPTW18} mitigate these unstructured memory accesses with an I/O-efficient implementation executing steps (i)~to~(iii) in batches.
By chosing a batch size of $\Oh{m}$ swaps, the previously $\Oh{m}$ unstructured I/Os can be transformed into a constant number of scans over the edge list;
the resulting pipeline triggers $\Oh{\sort(m)}$~I/Os \whp.
Its implementation is faster than~\cite{DBLP:journals/corr/abs-cs-0502085} even for instances still fitting into main memory, and scales well for graphs exceeding this threshold.

Bhuiyan~\etal~\cite{DBLP:conf/icpp/BhuiyanCKM14} propose a distributed approach for $\nproc$~processors.
\indDistributedMemory%
Each processor~$P_i$ is assigned approximately $m / \nproc$ edges~$E_i$, and generates swaps by selecting two random edges: one local edge $e_1 \in E_i$ and second (not necessarily local) edge $e_2 \in E$.
If edge~$e_2$ is stored on a different computer (\ie $e_2 \in E_j$ with $i \ne j$), processor~$P_i$ sends a message to the remote host~$P_j$ which then executes the swap.
In general, the algorithm uses additional messages to avoid multi-edges.
The implementation performs $\num{1.15e11}$ edge swaps on a network with $\num{1e10}$ edges in \SI{3}{h} using $\nproc=1024$~processors; \cite{DBLP:journals/jea/HamannMPTW18} carry out the same experiment on a single machine ($\nproc = 8$) with a slow-down of \num{8.3}.

	\subsection{Curveball and Global Curveball}
	\label{subsec:rand_curveball}
\iftrue
\begin{figure}
	\begin{center}
\begin{tikzpicture}
	\def\baseA{0.75}
	\def\baseB{0.1}
	\def\h{0.4}
	\def\adj#1{N_{#1}}

	\node[anchor=east, inner sep=0] at (0.2, \baseA + \h / 2) {$\adj{u}\colon$};
	\node[anchor=east, inner sep=0] at (0.2, \baseB + \h / 2) {$\adj{v}\colon$};
	
	\filldraw[fillColorAlt] (0.25, \baseA + 0.03) to ++(2, 0) to ++(0, \h) to ++ (-2, 0);
	\filldraw[fillColorAlt] (0.25, \baseB + 0.03) to ++(2, 0) to ++(0, \h) to ++ (-2, 0);

	\filldraw[fillColor] (2.5, \baseA + 0.03) to ++(4.25, 0) to ++(0, \h) to ++ (-4.25, 0);
	\filldraw[fillColor] (2.5, \baseB + 0.03) to ++(2.75, 0) to ++(0, \h) to ++ (-2.75, 0);

	\path[draw, thick] (0.25 + 2, \baseA + 0.03) to ++(0, \h);
	\path[draw, thick] (0.25 + 2, \baseB + 0.03) to ++(0, \h);
	
	\path[draw, thick] (0.25 + 2.25, \baseA + 0.03) to ++(0, \h);
	\path[draw, thick] (0.25 + 2.25, \baseB + 0.03) to ++(0, \h);

	\path[draw] (0.25, \baseA + 0.03) to ++(6.5, 0) to ++(0, \h) to ++ (-6.5, 0) to ++(0, -\h);
	\path[draw] (0.25, \baseB + 0.03) to ++(5, 0) to ++(0, \h) to ++ (-5, 0) to ++(0, -\h);

	\foreach \x in {0, ..., 25} {\path[draw] (0.25 + \x / 4, \baseA + 0.03) to ++(0, \h);}
	\foreach \x in {0, ..., 19} {\path[draw] (0.25 + \x / 4, \baseB + 0.03) to ++(0, \h);}
	
	\node at (2.5 - 0.125, \baseA + \h / 2) {\footnotesize $v$};
	\node at (2.5 - 0.125, \baseB + \h / 2) {\footnotesize $u$};
	
	\path[draw, ultra thick, <->, bend right]
		(4.75, \baseB + \h/2)
		to node[right, pos=0.5] {\ shuffle}
		(6, \baseA + \h / 2);
	
	\node[anchor=south, inner sep=0, align=center, font=\small] at (1.25, -1) {$\adj{u} \cap \adj{v}$\\ \phantom{j} common \phantom{j} };
	
	\node[anchor=south, inner sep=0, align=center, font=\small] at (4.75, -1) {$[\adj{u} {\cup} \adj{v}] \setminus [\adj{u} {\cap} \adj{v}] \setminus \{u{,}v\}$\\disjoint neighbors};
\end{tikzpicture}
 	\end{center}

	\caption{%
		A curveball trade.
		The neighbors of two randomly selected nodes~$u$, $v$ are shuffled.
		Common neighbors and the edge $\{u, v\}$ must not be traded for the graph to remain simple.
	}
	\label{fig:curveball_trade}
\end{figure}
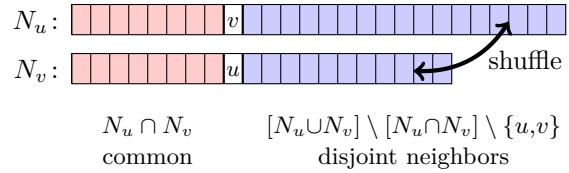
\fi

\indDegreeSequence%
\modLongCB (\modCB)~\cite{strona2014curveball, CarstensPhd} is structurally similar to \modLongES as it randomizes a graph by executing a sequence of local modifications, so-called \emph{trades}.
It is available for directed\footnote{%
	\indSelfLoops%
	If self-loops are disallowed, directed triangles cannot be reorientated by \modLongCB.
	Several preprocessing steps have been considered to lift this restriction (\eg the linear time algorithm~\cite{carstens1803.02624}).%
} and simple undirected graphs.
\indGraphDirected%
Each trade selects two nodes $u \ne v$ uniformly at random, and shuffles their neighborhoods (see \Cref{fig:curveball_trade}).
To this end, the set of disjoint neighbors $(\adj(u) {\cup} \adj(v)) \setminus (\adj(u) {\cap} \adj(v)) \setminus \{u,v\}$ is identified, and randomly redistributed between the nodes~$u$ and~$v$ while keeping their degrees unchanged.
Compared to the \modLongES Markov chain, the basic steps are more complex but at the same time inflict more changes; hence empirically fewer steps are required~\cite{Carstens2016}.
Additionally, \modCB trades increase data-locality~\cite{DBLP:conf/esa/CarstensH0PTW18}.

Carstens~\etal~\cite{Carstens2016} (extended by~\cite{DBLP:conf/esa/CarstensH0PTW18}) propose \modLongGCB (\modGCB) which further reduces data-dependencies.
A global trade is a super-step in the Markov Chain, and consists of $\floor{n/2}$ single trades targeting all nodes of the graph (if~$n$ is odd, a random node remains unselected).
The underlying Markov chain still converges towards a uniform sample, and in practice shows fast mixing times even for skewed degree sequences.
\indIOEfficient%
Carstens~\etal~\cite{DBLP:conf/esa/CarstensH0PTW18} introduce an I/O-efficient parallel algorithm that exploits the increased regularity of \modGCB's trading patterns. \indEndDegreeSequence%
\section{Block Models}
\label{sec:block_models}
\indCommunityDetection\indClustering%
The goal of \emph{community detection} (also known as \emph{graph clustering}) is to identify regions of a graph that are internally densely connected, but only sparsely connected to their outsides.
Such communities may be disjoint or overlapping.
\emph{Disjoint communities} partition the set of nodes of a graph into disjoint subsets.
For \emph{overlapping communities}, each node may belong to more than one community.
Numerous measures and algorithms have been proposed to formalize the fuzzy concept of a community and how to detect them (see~\cite{FORTUNATO20161} for a survey).

Many observed graphs (\eg derived from biological systems or social networks) have a significant community structure~\cite{Girvan02}.
Therefore, it is natural to also consider generators that explicitly generate such structure.
\indPlantedCommunity%
For the evaluation of community detection algorithms, synthetic benchmark graphs with planted communities are useful;
their outputs' consist not only of the graphs produced but also includes an assignment of nodes to communities.

In the following, we introduce the traditional \modLongSBM that is also used to theoretically analyze community detection algorithms.
Subsequently, we consider the commonly used \modLFR{} generator, and the more recently proposed \modCKB{} generator.
 
	\subsection{Stochastic Block Model}
	\label{subsec:mod_sbm}
The \modLongSBM~(\modSBM)~\cite{HOLLAND1983109, DBLP:journals/ftcit/Abbe18} (also \modPPM or \modIRG) is a versatile framework to model a fine-grained community structure.
There exists numerous extensions and generalizations (see~\cite{DBLP:journals/ans/LeeW19} for a recent survey).
Its base variant has the following model parameter: the number~$n$ of nodes, the number~$k$ of communities, a community probability distribution $p=(p_1, \ldots, p_k)$, and a symmetric matrix $P \in [0,1]^{k \times k}$.

The \modSBM yields an undirected simple graph~$G$ and a \emph{ground-truth} community assignment $\chi\colon V \rightarrow C$.
\indWeightedSampling%
To this end, each node independently selects its community $c_j \in C$ weighted with probability $p_j$ (see \cref{sss:weightedSampling}).
Subsequently, we introduce an edge between each node-pair $\{u, v\}$ independently with probability $P_{\chi(u),\chi(v)}$.

Observe that after assigning nodes to communities, algorithms for \modSBM and \modLongChungLu models are similar.
While in \modChungLu ``community blocks'' form endogenously if nodes share the same weights, these blocks are expressed explicitly in \modSBM.
Nevertheless, from an algorithmic point of view they can be dealt with similarly.
Thus, Alam~\etal~\cite{DBLP:conf/sc/AlamKVM16} and Moreno~\etal~\cite{MorenoPN18} directly generalize their respective \modChungLu generators to \modSBM.

	\subsection{R-MAT / Kronecker Graphs}
	\label{subsec:mod_rmat}
\begin{figure}
	\begin{subfigure}[b]{0.4\textwidth}
		\begin{center}
			\includegraphics[scale=0.8]{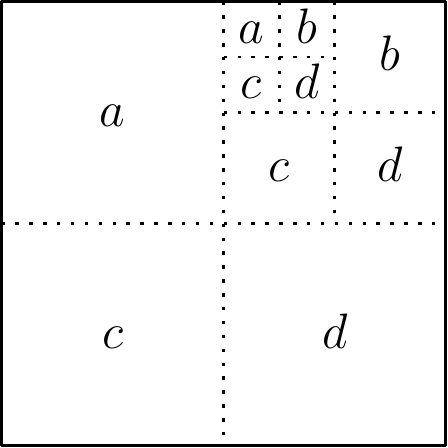}
		\end{center}
		\caption{\modRMAT schematic adjacency matrix}
		\label{fig:rmat_matrix}
	\end{subfigure}
	\hfill
	\begin{subfigure}[b]{0.5\textwidth}
		\begin{center}
			\begin{align*}
				\otimes&\colon \sR^{n\times m} \times \sR^{n' \times m'} \mapsto \sR^{(n \cdot n') \times (m \cdot m')}, \\[1em]
				\textbf{U} \otimes \textbf{V} &\Def \begin{bmatrix}
														u_{1,1}\textbf{V} & u_{1,2}\textbf{V} & \cdots & u_{1,m}\textbf{V}           \\[0.3em]
														u_{2,1}\textbf{V} & u_{2,2}\textbf{V} & \cdots & u_{2,m}\textbf{V}           \\[0.3em]
														\vdots & \vdots & \ddots & \vdots           \\[0.3em]
														u_{n,1}\textbf{V} & u_{n,2}\textbf{V} & \cdots & u_{n,m}\textbf{V}           \\[0.3em]
				\end{bmatrix}
			\end{align*}
		\end{center}
		\caption{The Kronecker product}
		\label{fig:kronecker_product}
	\end{subfigure}

	\caption[RMAT's recursive adjacency matrix]{
		(\ref{fig:rmat_matrix}) For each edge, \modRMAT recursively subdivides the adjacency matrix into quadrants, and each time selects one of them with probabilities $a$, $b$, $c$, and~$d$ respectively.
		The recursion terminates when a sub-matrix of size $1\times1$ is reached.
		(\ref{fig:kronecker_product}): Definition of the Kronecker product used to generalize this partitioning.
	}
\end{figure}

\modKroneckers~\cite{leskovec2005realistic} are a family of self-similar graphs that are based on the recursive application of Kronecker products (as defined in \Cref{fig:kronecker_product}) to the adjacency matrix of an initial seed graph.
\indPowerLaw\indDiameter%
These graphs obey static graph patterns, such as a power law degree distribution and a small diameter~\cite{leskovec2005realistic}.
\indGraphDensity%
Additionally, as these networks grow, their density increases.
To be more specific, the number of nodes and edges obey a densification power law, \ie $m$ is proportional to $n^\alpha$ for some $\alpha > 1$.
\indClustering\indDiameter%
The simplicity of \modKroneckers allows for the tractable analysis of various of these properties including the graph diameter, clustering coefficient and degree distribution~\cite{leskovec2010kronecker,mahdian2007stochastic}.
However, due to the discrete nature of the generation method \emph{staircase effects} can be observed in some of these quantities.

To alleviate this issue, Leskovec~\etal~\cite{leskovec2005realistic,leskovec2010kronecker} introduce \modStocKroneckers.
Instead of using an adjacency matrix, these graphs start with a probability matrix $\textbf{U}$ as their seed.
The seed matrix $\textbf{U}$ can be obtained from a \emph{deterministic} seed graph by replacing $0$-entries with $\alpha$ and $1$-entries with $\beta$ where $0 \leq \alpha \leq \beta \leq 1$ are model parameters.
Then, an entry $u_{ij} \in \textbf{U}$ corresponds to the probability that an edge between the vertices~$i$ and~$j$ is present.
The model then computes $\textbf{U}^k$ ($k$\textsuperscript{th} power of Kronecker products), and samples edges using probabilities prescribed in $\textbf{U}^k$.

In order to generate \modStocKroneckers that appear similar to a given graph~$G$,  Leskovec~\etal~\cite{leskovec2010kronecker} introduce a fast and scalable fitting algorithm called \algKronFit.
The algorithm avoids committing to specific metrics (\eg shape of degree distribution) by directly matching the adjacency matrix of~$G$ and the generated graph.
\algKronFit achieves a linear running time by exploiting the structure of Kronecker products and using statistical simulation techniques.

A special case of \modStocKroneckers is the \modLongRMAT (\modRMAT) by Chakrabarti~\etal~\cite{DBLP:journals/csur/ChakrabartiF06}.
This model is well-known for its usage in the popular Graph~500 benchmark.\footnote{%
	The benchmark additionally requires the generator to relabel all nodes based on a random permutation of node indices to avoid information leaking that could be exploited by algorithms tailor-made for the benchmark.
	\indRandomPermutation%
	As relabelling is straight-forward, we omit its discussion here.
}

A graph with~$n$ vertices and~$m$ edges is built by sampling each of the~$m$ edges independently.
To generate a single edge, \modRMAT partitions the adjacency matrix recursively into four equal-sized quadrants (see \Cref{fig:rmat_matrix}).
The quadrants are weighted with probabilities of $a$, $b$, $c$, and~$d$ respectively where $a + b +c + d= 1$.
The model then randomly selects one of the partitions to recurse on.
It continues until it reaches the base case of a $1 \times 1$ partition corresponding to the sampled edge.

Some further noise can be added at each step of the recursion to smooth out the degree distribution of the output graph.
This can be done by computing a level-dependent uniform random number during the recursion, and slightly modifying the initial probabilities for each quadrant~\cite{DBLP:journals/corr/abs-1102-5046}.
\indGraphDirected%
Note that this process produces a directed graph that may have multi-edges (which are typically converted into single edges).
Furthermore, directed graphs can be made undirected by setting $b=c$, and removing all matrix entries above the main diagonal and copying the lower half (\cf \cref{subsec:special_graphs_directed}).

The time complexity of the initially proposed \modRMAT generator~\cite{DBLP:journals/csur/ChakrabartiF06} is $\Oh{m \log n}$ since recursion has to be repeated for each edge.
Furthermore, this process is embarrassingly parallel, \ie edges can be generated independently of one another.
\indDistributedMemory%
This leads to a distributed memory algorithm with a runtime of $\Oh{[m/P] \log n}$.
Finally, once can easily generalize the resulting algorithm to generate \modStocKroneckers.

\indWeightedSampling%
The time for generating an edge can be reduced from logarithmic to constant using bit parallelism~\cite{HubSan19b}.
The algorithm precomputes sequences of a logarithmic number of decisions together with their probability.
These sequences can be sampled in constant time using the alias table data structure outlined in \cref{sss:weightedSampling}.
Thus, generating an edge just amounts to concatenating a constant number of sampled sequences of decisions.
 	
	\subsection{LFR}
	\label{subsec:mod_lfr}
\indGraphDirected\indGraphWeighted\indCommunityDetection%
The \modLFR{} benchmark has been first introduced for unweighted, undirected graphs~\cite{Lancichinetti2008} and later extended to directed and weighted graphs~\cite{Lancichinetti2009}.
\indPowerLaw%
The node degrees are drawn from a power law distribution with user-supplied parameters;
community sizes are drawn from an independent power law distribution.
A mixing parameter $\mu$ determines the fraction of neighbors of every node~$u$ that are not part of $u$'s communities.
The remainder of a node's degree, its internal degree, is divided evenly among all communities it belongs to.
Nodes are assigned to communities at random such that each node's internal degree is smaller than the size of the community as otherwise not enough neighbors can be found.
This is done using a bipartite version of \modLongESRef{} with additional rewiring steps to satisfy these degree constraints.
\indDegreeSequence%
For each community, \modLFR generates a graph with the given degree sequence using the \modLongFDSMRef.
It analogously samples the global \emph{inter-community} graph with the remaining degrees.
Additional \modES steps are used to rewire edges such that no edges of the global graph are within a community and to rewire edges that are part of multiple communities.

The authors provide sequential implementations of the different variants of the LFR benchmark~\cite{Lancichinetti2008,Lancichinetti2009}.
\indIOEfficient%
Hamann~\etal~\cite{DBLP:journals/corr/HamannMPTW17, DBLP:journals/jea/HamannMPTW18} propose an external memory algorithm that uses I/O-efficient \modES and a streaming implementation of \algHavelHakimi to generate graphs.
Rewiring steps are also implemented based on \modES{}.
If the number of communities is smaller than~$M$, the assignment can be solved in a semi-external algorithm in time $\Oh{\scan(n)}$ I/Os, otherwise I/O-efficient sampling is used that needs $\Oh{\sort(n)}$ I/Os.
The implementation of this external memory algorithm processes independent communities in parallel and outperforms the original implementation even for graphs still fitting into the internal memory.
Additional parallelism can be exposed by replacing \modES with \modLongGCB.
 	
	\subsection{CKB}
	\label{subsec:mod_ckb}
The \modCKB model~\cite{Chykhradze2014} is based on the \modLongAGM (\modAGM)~\cite{DBLP:journals/tist/YangL14}.
\modAGM encodes the assignment of nodes to communities using a bipartite graph.
Every community is similar to a $\Gnp$ graph with an individual edge probability $p$.
As a result, nodes sharing multiple communities have a higher probability of being connected.
An additional $\varepsilon$-community with a small edge probability $\varepsilon$ consisting of all nodes is added~\cite{DBLP:conf/icdm/YangL12a} to allow edges between any pair of nodes.
Yang and Leskovec show that this model captures many properties of real-world networks~\cite{DBLP:journals/tist/YangL14}.

The main goal of \modAGM is not graph generation but rather fitting this model to a real graph in order to detect communities.
Consequently, \modAGM provides no synthetic bipartite node-community graph, but instead uses structures taken from the observed graphs.
The \modCKB model mitigates this issue, and gives the means to randomly sample the necessary parametrization for \modAGM.
\indPowerLaw%
The number of communities a node is part of, as well as community sizes follow power law distributions.
The edge probability of a community $c_k$ is given by $p_{c_k} \Def \alpha / x_{c_k}^{\gamma}$, where $x_{c_k}$ is the number of nodes in $c_k$ and $0 < \gamma < 1$, $\alpha > 0$ are parameters.

The original implementation of \modCKB uses Apache Spark~\cite{DBLP:journals/cacm/ZahariaXWDADMRV16}.
\indDegreeSequence%
It first generates the two degree sequences for nodes and communities on a master node, and then sends them to the other worker nodes.
The number of edges for each community is sampled from a binomial distribution.
In order to compensate for multi-edges and self-loops, \modCKB calculates their expected number and samples more edges.
The generator then uses the \modLongECM to sample the assignment of nodes to communities, and to generate the individual community graphs.
\indDistributedMemory%
For both steps, each worker emits edge subset of similar sizes, which are then distributed and merged using the Hadoop Distributed File System (HDFS).
On Amazon EC2 with 100 m1.large instances (two virtual cores, \SI{7.5}{\giga\byte} RAM each, which might be a 2010 Intel Nehalem server processor~\cite{Kennedy13}), the authors are able to generate a graph with one billion nodes in less than 2 hours~\cite{Chykhradze2014}.
 
\section{Graph replication}
\label{sec:graph_replication}
The practical goal of \textsl{graph replication} methods is to generate graphs that are similar to one or more reference graphs.
\indDegreeSequence%
We already discussed a simple variant in \cref{sec:graph-randomization}, where the goal is to sample graphs with matching degree sequences.
However, depending on the application, other or additional properties are of importance.
Graph replication is typically used for tasks, such as algorithm testing, performance evaluation, benchmarking, and assisting decision makers in various domains including (but not limited to) engineering, software design, epidemiology, and viral marketing.

In such domain dependent tasks, a similarity between the synthetic and original graphs is often not well-defined.
Thus the designers of graph generators have to be careful when quantifying the \emph{realism} of a synthetic graph.
In some cases, such realism is measured in specific structural properties of paths, loops, special forms of network backbones at various resolutions, and connectivity between node clusters.
Hence traditional general properties (\cf \cref{sec:graphprop}) seem insufficient to generate a useful synthetic network.
For example and as discussed in \cref{subsec:graph_repl_multiscale}, the preserved properties may include the second shortest path length~\cite{DBLP:conf/fusion/GutfraindSM15}, or the graph Laplacian~\cite{shine2019generative}.
 
	\subsection{BTER}
	\label{subsec:mod_bter}
\indDegreeSequence\indClustering%
The \modLongBTER (\modBTER) model generates graphs approximately following a prescribed degree sequence and clustering coefficient per degree~\cite{DBLP:journals/siamsc/KoldaPPS14}.\footnote{%
    The original formulation~\cite{PhysRevE.85.056109} by a subset of the authors did not explicitly consider the clustering coefficient.%
}
Its input parameters are $\seq{d_i}{i=1}{d_\text{max}}$ and $\seq{c_i}{i=1}{d_\text{max}}$ where $d_i$ specifies the number $d_i$ of nodes with degree~$i$ and $c_i$ requests that all nodes of degree~$i$ to have a clustering coefficient of $c_i$.
Both quantities are only realized in expectation.

\modBTER generates graphs in two steps.
In the first phase, it constructs \emph{homogeneous affinity blocks} by grouping together $d{+}1$ nodes of degree~$d$ in a greedy fashion.
The remaining unassigned nodes form so-called \emph{heterogeneous affinity blocks} of mixed degrees (treatment omitted here, refer to~\cite{DBLP:journals/siamsc/KoldaPPS14}).
Each homogeneous affinity block is materialized as a $G(d+1, p)$ graph where~$p$ is chosen to match the requested clustering coefficient~$c_d$ in expectation.
Unless an affinity block forms a clique, the prescribed degrees are not met and additional edges need to be added in a second phase.
To this end, \modBTER computes for each node $v \in V$ its excess degree $e_v$ as the difference between its requested degree and the degree expected from phase~1.
It then supplements the missing edges with a \modChungLu graph on the degree sequence $\degseq = \seq{e_v}{v}{}$.
Ultimately the resulting graph is constructed by merging the subgraphs from both phases.

Kolda~\etal propose a parallel sampling algorithm~\cite{DBLP:journals/siamsc/KoldaPPS14}:
each worker independently adds edges by first randomly selecting for which phase the edge is introduced.
Then the node pair is drawn from the appropriate distribution.
This process almost surely generates duplicate edges.
To avoid this bias, more edges are generated to account for the expected losses during de-duplication.
Similarly, the number of nodes with degree~1 is scaled by a correction factor~$\beta>1$ to compensate that approximately 36\,\% of them remain singletons while 28\,\% receive at least two neighbors~\cite{DBLP:conf/nsw/DurakKPS13}.

Based on their efficient sequential and parallel implementations of \modChungLu, Alam~\etal~\cite{DBLP:conf/sc/AlamKVM16} also present efficient sequential and parallel implementations of \modBTER.
They are able to generate a graph with 131 million nodes and 4.6 billion edges in 210 seconds sequentially and just 0.37 seconds using 1024 processors, \ie they achieve a speedup of about 572.

\modGBTER~\cite{DBLP:journals/snam/BridgesCFLS16} (\modLongGBTER) generalizes \modBTER by allowing the user to supply the groups of the nodes instead of automatically grouping nodes by degree.
This enables the generation of a graph with a prescribed community structure.
The user can also supply a density per group, replacing the automatically assigned density to match a certain clustering coefficient.
However, there is no support for matching a certain clustering coefficient.
\modEGBTER~\cite{DBLP:conf/asunam/El-DagharLB18} (\modLongEGBTER) adds support for clustering coefficients in addition to the user-supplied groups.
\modEGBTER takes a community assignment $\chi\colon V \rightarrow C$, the groups, the expected within-community degree of each node $\seq{d_i}{i=1}{n}$, the expected global degree of each node $\seq{D_i}{i=1}{n}$, with $d_i \leq D_i$, and the within-community clustering coefficient distribution $\seq{c_i}{i=1}{d_m}$.
For the generation process, the authors basically run \modBTER for every community separately and then generate an additional \modChungLu graph for the remaining degrees.
Thus, scalable implementations of \modBTER can be easily adapted for \modEGBTER.

\indCommunityDetection%
\modABTER~\cite{slota19} (\modLongABTER) uses a specially configured version of \modBTER to generate benchmark graphs for community detection similar to the \modLFR benchmark.
At the same time, \modABTER replicates the degree and clustering coefficient distribution of a given graph similarly to \modBTER.
Using a heuristic scaling mechanism, \modABTER slightly adapts these distributions to match a prescribed average and maximum degree and clustering coefficient on a possibly larger graph.
\modABTER takes a mixing parameter that, like in the \modLFR model, denotes the fraction of inter-cluster edges.
Using a linear program, \modABTER further adjusts the degree and clustering coefficient distributions such that the resulting graph matches the mixing parameter while keeping the adjustments minimal.
The assignment to affinity blocks, and the edge generation closely follows the \modBTER model.
Using the parallel edge-skipping technique~\cite{DBLP:conf/sc/AlamKVM16}, the actual intra- and inter-community edges are generated efficiently in a parallel, distributed implementation.
\indDistributedMemory%
The implementation is based on MPI and OpenMP.
On 512 compute nodes where each node has \SI{128}{\giga\byte} RAM and two marvel ThunderX2 ARM processors with 28~cores per processor, \modABTER generates a graph with 925~billion edges and 4.6~billion nodes in 76~seconds.
While the resulting graphs do not perfectly match the original \modLFR model, they show that community detection algorithms show similar behavior.
Further, they show that both the degree and clustering coefficient distributions as well as the mixing parameter are as desired. 	
	\subsection{Darwini}
	\label{subsec:mod_darwini}
\indClustering%
The \modDarwini{}~\cite{DBLP:conf/icdcs/EdunovLWCK18} generator takes both a degree distribution, and a distribution of clustering coefficients per degree as input.
Usually, such distributions would be obtained from an observed network.
Using this input, \modDarwini can generate a graph of a different scale that is similar to a real-world network from which the distributions have been used.
The authors show that it outperforms other approaches like \modBTER{} in terms of the reproduced metrics on the Facebook social graph.

\modDarwini{} first samples for every node a degree, and then a clustering coefficient from the distribution of that degree.
The basic idea of \modDarwini{} is then to group nodes into buckets, where all nodes in a bucket have the same (or a similar) desired number of triangles in order to achieve their clustering coefficient.
For each bucket, a $\Gnp$ graph is generated.
Bucket sizes and probabilities~$p$ are chosen such that the expected number of triangles in each bucket matches the desired number while ensuring that no node in the bucket can exceed its target degree.
After this first step, each node has a residual degree left.
\modDarwini{} iteratively adds edges between buckets until it achieves the target degrees.
Roughly speaking, it is unlikely that these inter-bucket edges create new triangles.
Thus, this step allows to both reach target degrees while keeping the target clustering coefficients.
In each iteration, the algorithm attempts two strategies for creating edges.
Firstly, from each node, it attempts to create an edge to a random node, which succeeds if the other node has a non-zero residual degree.
Secondly, it shuffles the nodes into small groups and attempts adding edges between nodes within a group.
For edges within a group, the algorithm favors edges between nodes of similar degree.

The authors provide an open source implementation of \modDarwini{}\footnote{\url{https://issues.apache.org/jira/browse/GIRAPH-1043}} in Apache Giraph\footnote{\url{https://giraph.apache.org}}.
While the implementation carries out the assignment to buckets on a central compute node, it distributes the creation of edges within buckets while processing each bucket sequentially.
Globally uniformly distributed edges are generated in parallel while the random groups are again aggregated to generate the edges within the group.
\modDarwini introduces an additional layer of super-communities to generate graphs larger than the main memory of the compute cluster.
It is executed individually for each super-community, and then later edges between nodes in super-communities are generated by sampling potential neighbors and testing if they still want new neighbors.
The authors generated a scaled up version of the entire Facebook social graph with 3 trillion edges in 7 hours on a compute cluster of 200 machines each with \SI{256}{\giga\byte} RAM and 48 cores.

	\subsection{Multilevel generators}
	\label{subsec:graph_repl_multiscale}
Multilevel algorithms for computational optimization on graphs are well-known to be successful on such problems as partitioning~\cite{safro2015advanced}, linear arrangement~\cite{ron2011relaxation}, force-directed drawing~\cite{walshaw2000multilevel}, and vertex separator~\cite{hager2018multilevel}.
The main idea behind these algorithms is to create a hierarchy of increasingly coarse representations in which each next-coarser graph is structurally similar but smaller than the current level coarse representation.
While constructing a hierarchy, an optimization problem is solved for each coarse representation (from the coarsest level to finest) in a such way that a coarse level solution serves as an approximation for the next-finer level.
Here we describe how the multilevel algorithm design pattern can be used for graph generation.

The multilevel\footnote{%
    In the original paper, the authors used the term \emph{multiscale} to emphasize that the generator acts at multiple scales of a complex system coarseness.
    Here we replace it with the term \emph{multilevel} to avoid possible associations with scalable performance.%
}
graph generator \modMUSKETEER~\cite{DBLP:conf/fusion/GutfraindSM15} deems the following two observations important to generate realistic replica:
\indClustering\indDiameter%
Firstly, various graph properties ---while being different in absolute terms--- are as important on the coarse levels as on the original finest level.
For example, both clustering coefficient and diameter, can be illuminating at the coarse levels, when the finest scale information is aggregated after coarsening.
Secondly, applying a single node or edge editing operation (such as adding/removing a node or adding/removing/rewiring an edge) at a coarse level would be equivalent to applying many operations on the whole regions in a graph at the finest level.
On the one hand, this will contribute to the realism of a generated graph because the local structure will be preserved if changes are applied at the coarse levels.
Conversely, if the changes are applied only at the fine levels, the global structure of a graph will remain unchanged.
As a result, a user can control at what levels, editing operations should be applied thus controlling the desired deviation from the original graph at both local and global resolutions.

\modMUSKETEER follows the same coarsening scheme as in multilevel approaches for combinatorial optimization problems, such as partitioning and linear ordering~\cite{ron2011relaxation,safro2015advanced}.
However, it edits synthetic graph preserving the required properties at each level during the uncoarsening phase instead of optimizing some objective.
Throughout all levels, all editing operations are local.
For example, in~\cite{DBLP:conf/fusion/GutfraindSM15}, a distribution of the second shortest path length measured at each level during the coarsening is preserved when a new node is added and connected to other nodes.
The nearly linear complexity of this approach is comparable to that of other multilevel optimization algorithms.
Reported preserved metrics include  degrees, assortativity, eccentricity, clustering, betweenness centrality and harmonic mean distance centrality.

The multilevel approach can also generate planar graphs~\cite{chauhan2018multiscale} by applying linear time planarity test (for Kuratowski subgraphs~\cite{thomassen1981kuratowski}) followed by rejecting added edges that violate planarity at each level of coarseness.
\indGraphPlanar%
The graph editing at each level is required to only produce planar graphs.

Indirectly, the multilevel approach attempts to preserve the spectral properties of an original graph either at the user specified, or all levels of coarseness.
Conversely, Shine and Kempe propose \modSpectralGen~\cite{shine2019generative} to intentionally produce graphs with a similar (or matched) spectrum of the graph Laplacian.
The spectrum of the Laplacian encodes high-level connectivity information about the original graph, and can be controllably preserved within the multilevel frameworks using advanced coarsening schemes~\cite{chen2011algebraic}.

\modSpectralGen acts as follows:
Firstly, it generates a template matrix with a spectrum close to the original graph Laplacian using a randomly sampled orthonormal basis followed by a linear programming based fitting.
Since the result is not necessarily a correct graph Laplacian, \modSpectralGen subsequently rounds the matrix to a valid graph Laplacian using linear programming.
The approach is currently prohibitive expensive for large-scale graphs.
There is no theoretical guarantee that the generator samples uniformly from the set of graphs with approximately correct spectra.
The authors, however, experimentally demonstrate a significant variation in the generated graphs while preserving useful metrics including betweenness centrality and path length distribution.

One of the interesting future challenges is combining other generative approaches with the multilevel framework by applying them at different levels of coarseness.
For example, one may want to generate a graph with specific global geometry at the coarse levels while addressing local properties, such as clustering coefficient or degree distribution, at the fine levels only.
 
	\subsection[dK-Graphs]{$dK$-Graphs}
	\label{subsec:mod_dkgraphs}
Mahadevan~\etal~\cite{DBLP:conf/sigcomm/MahadevanKFV06} propose \moddKGraphs, a family of models parametrized by~$d$.
The parameter controls the level of details captured from the input graph:
\begin{itemize}
    \item For $d = 0$, only the average degree is kept.
    \item For $d = 1$, the degree distribution is reproduced analogously to \modFDSMRef.
    \item For $d = 2$, for every pair of degrees the number of connections between nodes of these degrees is preserved.
    \item For $d > 2$, for every possible induced subgraph with~$d$ nodes, the occurrences of a certain $d$-tuple of degrees is preserved.
    \item For $d = n$, the input graph is replicated exactly.
\end{itemize}

The authors report that they obtain graphs very similar to an internet topology graph supplied as input already for $d = 3$.
Given an input graph, they use a restricted version of \modLongES{} that only allows switches that preserve the considered distributions.
Given just the desired properties and no input graph, the authors consider edge-switches that bring them closer to the desired property.
They report no running times, but report that the algorithmic complexity increases sharply with increasing values of~$d$.
However, for $d = 2$, it should be possible to adapt efficient implementations for \modLongES{}.
For $d > 2$, it seems unlikely that such graphs could be generated efficiently as an increasing number of edge switches would need to be rejected and there are no obvious ways to select them directly. 
\section{Additional graph types}
\label{sec:special_graphs}
Applications often mandate special types of networks:
street networks, for instance, are \emph{directed} to account for lanes or one-way-streets, and may express the distance between two points as \emph{edge weights}.
In the following, we outline modifications of previously discussed models and generators to cater to such use cases.
For the sake of brevity, we focus on general techniques and selected examples rather than aiming for an exhaustive survey.
 
	\subsection{Directed graphs}
	\label{subsec:special_graphs_directed}
\indDefGraphDirected%
Algebraically, a digraph is a binary $n \times n$ adjacency matrix $A$.
For an undirected graph, the matrix~$A$ is symmetric with an empty diagonal and, thus, fully determined by either its upper or lower triangle matrix.\footnote{%
	Many generators, \eg \modER (\cref{subsec:mod_erdos_renyi}), \modRMAT (\cref{subsec:mod_rmat}), or \modRHG (\cref{subsec:mod_hyperbolic_graphs}), explicitly exploit this fact by only sampling from one triangle matrix.
	The other triangle matrix then follows due to symmetry.%
}
Therefore, the set of all digraphs with~$n$ nodes is exponentially larger than its undirected counterpart.
As a direct consequence, random graphs models need to be extended to properly deal with digraphs.

In the simplest case the two edges $(u, v)$ and $(v, u)$ are treated independently (\eg as in the directed variants of \modLongER and \modGilbert models discussed in \cref{subsec:mod_erdos_renyi}).
\indGraphDirected\indPreferentialAttachment%
Bollob\`as~\etal propose a directed preferential-attachment model (see \cref{subsec:mod_preferential_attachment}).
In each iteration, their generative model randomly selects one of three actions:
add an edge with a new node as tail, add an edge with a new node as head, or connect two existing nodes.
When selecting a random edge source, node~$v$ is chosen with probability linear in $\degree_\text{out}(v)$.
Analogously, edge heads are selected based on the nodes' in-degrees.
Most of the algorithms discussed in \cref{subsec:mod_preferential_attachment} for the undirected case carry over by treating edge heads and tails separately.
The main observation is that node~$u$ with in-degree $d_u^\text{in}$ appears exactly $d_u^\text{in}$ times as an edge head, \ie at odd position in the edge array~$E$ (and analogously for out-degrees).
Hence, random edge tails are copied from uniformly selected even indices, and edge heads from uniformly selected odd positions.

Bogu{\~{n}}{\'{a}}~\etal~\cite{Boguna2014} propose a directed random graph model influenced by \modRHG.
It captures causality relations on cosmological scales, \ie where the speed-of-light limits what is observable.
Nodes have a random position in space and time, and a directed edge $(u, v)$ is added if~$u$ can be observed by $v$.

The notion of a degree sequence $\degseq = \seq{d_i}{i=1}{n}$ can also be extended to digraphs by replacing each degree by a 2-tuple $\degseq_\text{directed} = \seq{(d_i^\text{in}, d_i^\text{out})}{i=1}{n}$.
Then, the \modLongCM uses two independent urns---one for edge heads, one for edge tails.
Similarly, the \modFDSM model is available for digraphs.
\indDegreeSequence%
It is the building block for a directed variant of the \modLFR benchmark which otherwise draws and processes the in- and out-degree sequences independently.
The necessary modifications to the deterministic \algHavelHakimi generator, \modLongES, and even \modLongCB are mostly obvious.
Here, one notable exception is that the Markov chains of \modES and \modCB are not irreducible on the ensemble of simple digraphs because the direction of a directed 3-cycle cannot be reversed~\cite{CarstensPhd}.
One solution is a linear-time preprocessing step by Berger and Carstens~\cite{carstens1803.02624}.
 
	\subsection{Weighted graphs}
	\label{subsec:special_graphs_weighted}
We model \emph{weighted networks} by augmenting a graph $G = (V, E)$ with weights~$w\colon E {\to} \sR$, and define the \emph{strength} $s(u)$ of node~$u$ as the sum of weights of all edges incident to~$u$.
The semantics of \emph{edge weights}~$w$ are domain-specific;
examples range from energy flows in food webs, over capacities in transportation networks, to costs or distances in street maps.
For positive integer weights, $w(e)$ can also be interpreted as the multiplicity of edge~$e$---this is especially sensible when modeling capacities along edges~\cite{PhysRevE.70.056131}.
Here, a multiplicity $w(e) = 0$ encodes the absence of edge~$e$.

\indDefGraphWeighted%
The arguably simplest way to obtain a weighted graph is to assign random edge weights drawn independently and identically distributed to a random graph.
This is a common theme in both theoretical (\eg shortest path, minimum spanning trees, or first passage percolation) and empirical studies (\eg \modLFR variant~\cite{Lancichinetti2009}).

The \modLongWRG model (\modWRG)~\cite{garlaschelli2009weighted} is a maximum entropy model proposed as the weighted variant of $\Gnp$ and implements this idea.
The distribution $\modWRG(n, p')$ is defined over $\gGn$ and assigns each potential edge~$e$ the multiplicity $w(e)$ drawn from the geometric distribution $\geom{p'}$.
Since any edge~$e$ exists with $\prob{w(e) > 0} = 1 - p'$, the \modWRG model is topologically equivalent to \Gnp with $p = 1 {-} p'$.
The generators discussed in \cref{subsec:mod_erdos_renyi} therefore carry over with little modifications.

In real networks, however, the assumption of an independence between topology and edge weights does not hold.
Serrano and Bogu\~n\'a, for instance, observe non-trivial correlations between a node's degree and strength~\cite{serrano2005weighted}.
\indDegreeSequence\indPowerLaw%
The authors then analyze the \modLongCMRef where the number of balls for each node follows a power law degree sequence with small exponent $|\gamma| \ll 3$.
They derive the distribution of degrees and strength in the limit of $n \to \infty$, show that both follow different power law distributions, and demonstrate non-trivial correlations.
Instances of this model can be sampled by counting\footnote{%
	Commonly studied as the word count problem, it can be efficiently computed in most practical machine models.%
}
the multiplicities emitted by a standard \modCM generator.
The \modLongSBMRef can be extended in the same spirit (\eg~\cite{kurihara2006frequency}).

Britton~\etal~\cite{britton2011weighted} propose another weighted \modCM variant.
It features a second family of distributions assigning each ball a weight in addition to the number of balls for each node.
It then only pairs balls of equal weights;
if the number of balls with a given weight is odd, a random ball is dropped.
This model can be efficiently implemented, by treating each weight class as an independent urn.

\indGraphSpatial%
In spatial graphs, a distance function induced by the underlying geometry is a natural choice for edge weights.
For example, one could augment \modRGGs with edge weights that give the Euclidean distance of the connected points.

Other, application-specific solutions have been studied.
Hyun-Joo~\etal~\cite{kim2002weighted}, for instance, consider a finance network where each node corresponds to a market player.
The agents are augmented with latent variables indicating their performance.
Then weights are computed based on latent variables of their endpoints.
 	
	\subsection{Connected graphs}
	\label{subsec:special_graphs_connected}
Few of the previously discussed models guarantee that the generated graphs are connected.
\indPreferentialAttachment%
Preferential attachment models are a note-worthy exception as they stay connected if the provided seed graph is.
Additional well-known models with efficient generators include \modLongRDT and the Watts-Strogatz model~\cite{watts1998collective}.
Another source of very sparse and connected graphs are random regular graphs (see \cref{subsec:special_graphs_regular}).

There are three general techniques to obtain a connected graph from an existing model.
In the following we consider $\Gnp$ graphs.
They undergo phase transitions as~$p$ is increased~\cite{erdos1960evolution}.
In the extreme of $np > (1+\varepsilon)\ln n$, the graphs are connected with high probability.
In such cases \emph{rejection sampling} leads to efficient generators.
\indRejectionSampling%
Here, one generates graphs $G_1, G_2, \ldots$ and returns the first~$G_i$ that is connected.

Yet, $\Gnp$ almost surely yields isolated nodes for $np < (1-\varepsilon)\ln n$.
This renders rejection sampling an unsuitable choice in this parameter region.
Another approach is to identify the largest connected component~$C$ of a randomly drawn network and to remove all nodes and edges not contained (or incident with) $C$.
This selection process can introduce biases and must not preserve properties of the original model.
Additionally, the number $n'$ of nodes of the resulting graph is a random variable.

In order to be efficient, $n'$ should be sufficiently close to the number~$n$ of nodes emitted by the original random graph model.
This is the case for models featuring a large giant component\footnote{%
	For some positive constant~$c$, a connect component~$C \subseteq V$ in $G=(V,E)$ is a \emph{giant component} if $c\card{C} > \card{V}$.}.
While there exist general characterizations for the presence of a giant component (\eg based on a graph's degree distribution~\cite{erdos1960evolution}), many random models have more precise predictions.
The threshold variant of \modRHG, for instance, has a giant component of size $\Omega(n^{1 - 2\alpha})$ \whp rendering filtering for small values of $\alpha$ efficient.
The same is true for $\Gnp$ if $np$ is sufficiently close to $\ln n$.

Inversely to filtering, one can introduce additional edges to ensure connectivity.
A simple oblivious method is to add a random spanning tree (\eg~\cite{DBLP:conf/iccS/RodionovC03}) to an arbitrary graph.
Alternatively, one can selectively introduce bridges to iteratively merge disjoint connected components.

The edges of a prescribed connected graph can be shuffled using \modLongES.
Here, the Markov Chain is the normalized, induced subgraph of the original Markov Chain discussed in \cref{subsec:rand_edge_switching}, where the state space is reduced to connected graphs.
\indDegreeSequence%
Mihail~\etal~\cite{Mihail2003}  show that this modified version still eventually leads to a uniform stationary distribution and discuss characterizations of compatible degree sequences.
Viger and Latapy~\cite{DBLP:journals/compnet/VigerL16} further engineer the \modES for connected graphs.
 	
	\subsection{Regular graphs}
	\label{subsec:special_graphs_regular}
\begin{figure}
	\begin{subfigure}[b]{21em}%
\begin{tikzpicture}[
	vertex/.style={draw, circle, inner sep=0, minimum width=1.5em, minimum height=1.5em},
	edge/.style={draw, thick}
]
	\foreach \i [count=\x] in {0, ..., 4} {
		\node[vertex] (u\x) at (360/5*\i+90: 3em) {$v_\x$};
		\node[vertex, xshift=12em] (v\x) at (360/5*\i+90: 3em) {$v_\x$};
	}

	\draw[edge] (u4) to (u5);
	\draw[edge] (u2) to (u3);
	\draw[edge, out=240,in=300,looseness=8, ultra thick] (u1) to (u1);
	
	\draw[edge] (v1) to (v2);
	\draw[edge] (v1) to (v5);
	\draw[edge] (v3) to (v4);
	
	\path[draw, thick, bend left, ->, black!50] (4em, 2em) to node[above] {forward} (8em, 2em);
	\path[draw, thick, bend left, ->, black!50] (8em, -1em) to node[below] {backward} (4em, -1em);	
\end{tikzpicture} %
		\caption{$l$-switch.}
	\end{subfigure}\hfill
	\begin{subfigure}[b]{17.5em}%
\begin{tikzpicture}[
	vertex/.style={draw, circle, inner sep=0, minimum width=1.5em, minimum height=1.5em},
	edge/.style={draw, thick}
]
	\def\sepy{-2.8em}

	\foreach \i in {1, 2, 3} {
		\node[vertex] (a\i) at (0em, \i*\sepy)   {$u_\i$};
		\node[vertex] (b\i) at (3em, \i*\sepy) {$v_\i$};

		\node[vertex] (u\i) at (12em, \i*\sepy)   {$u_\i$};
		\node[vertex] (v\i) at (15em, \i*\sepy) {$v_\i$};
	}

	\path[draw, thick, bend left, ->, black!50] (5.5em, 2*\sepy + 1.5em) to node[above] {forward} ++ (4em, 0);
	\path[draw, thick, bend right, <-, black!50] (5.5em, 2*\sepy - 1em) to node[below] {backward} ++ (4em, 0);	

	\path[edge] (a1) to (b1);
	\path[edge, bend left, ultra thick] (a2) to (b2);
	\path[edge, bend right, ultra thick] (a2) to (b2);
	\path[edge] (a3) to (b3);
	
	\path[edge] (u1) to (u2);
	\path[edge] (u2) to (u3);

	\path[edge] (v1) to (v2);
	\path[edge] (v2) to (v3);	
\end{tikzpicture} 		\caption{$d$-switch.}
	\end{subfigure}

	\caption{%
		The $l$-switch removes a single loop (at $v_1$);
		the $d$-switch removes a double edge (edge $\{u_2, v_2\}$).
		The remaining nodes are selected such that no new loop or double edge is created.
	}
	\label{fig:dl-switchings}
\end{figure}
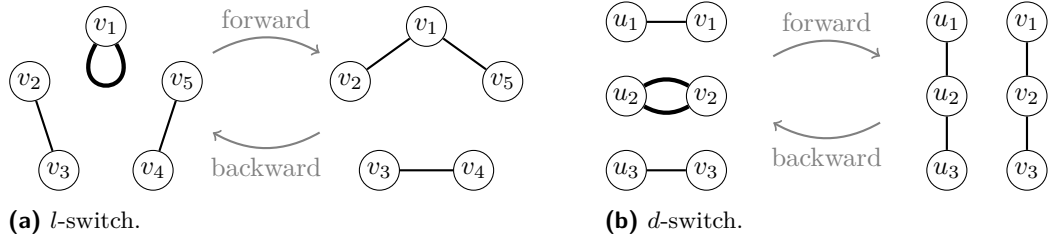

\indDefGraphRegular%
\indDefDegreeSequenceRegular%
A graph~$G$ is said to be \emph{$r$-regular} for $r \in \sNpos$ if all nodes have degree exactly~$r$~\cite{bollobas1985random, wormald1999models}.
This implies $r < n$ and that $n\cdot r$ is even.
In the following, we denote the set of all $r$-regular graphs with~$n$ nodes as $\gGnr$ and the uniform distribution over it as $\Gnr$.

Observe that $\gGnr \subset \gGnm$ since all graphs in $\gGnr$ contain $m = n\cdot r / 2$ edges.
However, regular graphs appear relatively infrequently~\cite{DBLP:journals/jct/BenderC78} with $|\gGnr| / |\gGnm| = \Theta[\exp(\frac{1-r^2}{4})]$ in the limit of ${n \to \infty}$.
As a result, $\gGnr$ exhibits substantially different properties compared to~$\gGnm$;
for instance, a random $G \in \Gnr$ for $r \ge 3$ is almost surely \emph{$r$-connected}\footnote{%
	A graph is $r$-connected if it remains connected after the removal of any $r{-}1$ nodes.%
}~\cite{bollobas1985random}.
In contrast, a uniform sample $G \in \Gnm$ is $1$-connected \whp only for an average degree of $ \Omega(\ln n)$ (\cf \cref{subsec:special_graphs_connected}).
This renders $\gGnr$ of particular interest to obtain random connected graphs of small constant degree~$r$.
Additionally, almost all $\gGnr$ with $r \ge 3$ are Hamiltonian~\cite{DBLP:journals/rsa/RobinsonW94}.

\indDegreeSequence%
To sample from $\Gnr$ all techniques discussed for the \modLongFDSMRef carry over since $\gGnr$ can be fully characterized by the degree sequence $\degseq = \seq{r}{i=1}{n}$.
\indRejectionSampling%
Most prominently, rejection sampling over the \modLongCMRef is efficient for small constant values of~$r$.

McKay and Wormald~\cite{DBLP:journals/jal/McKayW90} propose \algDeg, a more versatile scheme for $r = \Oh{n^{1/3}}$.
Rather than rejecting all non-simple graphs emitted by \modCM, the generator also accepts some multi-graphs.
The authors show that this selection yields a constant acceptance probability.
Then, \algDeg iteratively removes loops and double edges using the two switching types illustrated in \Cref{fig:dl-switchings} -- in each iteration one switch is selected uniformly at random to remove exactly one defect.
Since these switches cause a bias, the emitted graphs are not uniformly distributed on $\gGnr$ anymore.
To counteract this bias, each switch may be rejected with small probability.~\footnote{%
	The bias is because, in general, the number of multi-graphs with $\ell$ loops ($k$ double edges) does not match the number of multi-graphs with $\ell{-}1$ loops ($k{-}1$ double edges).
	Thus, there are fewer switches that decrease the number of loops (or double edges) than the inverse operation~\cite{DBLP:journals/jal/McKayW90}.
	As a countermeasure, the rejection probability is derived from the ratio of forward- and backward-switches.
	In fact, it is over-estimated as the exact computation is too expensive.
	Since each switch is selected uniformly at random and independently of the rejection step, the process eventually yield uniform samples from $\gGnr$---even if restarts occur more frequently than required.
}

The generator \algDeg was improved twice culminating in \algIncReg.
In a first step, Gao~\etal~\cite{DBLP:journals/siamcomp/GaoW17} introduce additional switches which do not decrease the number of defects but allow for tighter bounds on the rejection probability.
\algIncReg~\cite{DBLP:conf/focs/ArmanGW19} additionally allows triple edges in the multi-graph and adds new switches to remove them.
The authors further accelerate the switching process by simplifying the way nodes participating in a switch are sampled.
As a result, an illegal switch (\eg one introducing multi-edges) can be selected.
In this case the computation is restarted, effectively splitting the rejection step of the \algDeg into two simpler steps.
\algIncReg has an expected runtime of $\Oh{rn + r^4}$ for $r = o(\sqrt n)$ which is optimal for $r = \Oh{n ^ {1/3}}$.
 	
	\subsection{Threshold graphs}
Threshold graphs were introduced by Chvátal and Hammer~\cite{Chvatal77}.
Mahadev and Peled~\cite{mahadev1995thresholdgraphs} discuss numerous characterizations and applications of threshold graphs.
Just like their superclass of split graphs, they can be defined in terms of their degree sequence.
Alternatively, they can be obtained by iteratively adding nodes that are either connected to all previously added nodes (dominating nodes) or none of them (isolated nodes).
By randomizing the decision what kind of node to add, we immediately obtain a random graph generation algorithm that is linear in the size of the generated graph.
Generating~$n$ nodes such that dominating and isolated nodes are chosen with equal probability yields a uniform distribution on the set of unlabeled threshold graphs of~$n$ nodes~\cite{Holmes08}.
Due to the simplicity of this algorithm, it can be transferred to a distributed setting without communication.
\indDistributedMemory\indCommunicationFree%
For every node, we consistently flip a coin to determine if it is a dominating or isolated node.
If it is dominating, we emit edges for all node ids that are smaller than $u$.
To also generate edges to nodes with larger node ids, we need to repeat the decision for all nodes~$v$ with larger node id and emit an edge if~$v$ is a dominating node.
As the probability of dominating nodes is 0.5, this does not increase the running time in expectation. 
\section{Software packages}
\label{sec:software}
\iffalse
	\def\softwareTableMiniPageWidth{110mm}
	\def\softwareTableDescWidth{120mm}
\else
	\def\softwareTableMiniPageWidth{86mm}
	\def\softwareTableDescWidth{94mm}
\fi

\newcommand{\software}[5]{%
  #1 &%
  \begin{minipage}{\softwareTableMiniPageWidth}%
  	\url{#2} $\cdot$ #3%
  \end{minipage} &%
  #4&%
  \begin{minipage}{14mm}#5\end{minipage}\hspace{-6mm}\\%
}%
\begin{table}[bt]
 \renewcommand{\arraystretch}{0.8}
 \setlength{\tabcolsep}{3pt}
 \caption[List of publicly available implementations sorted by name of the toolkit]{
 	List of publicly available implementations sorted by name of the toolkit. \small
 	Abbrv.:
	\foreach \x in {BA, ER, ES, FDSM, RDT, RGG, RHG, SBM, WS}{
		\underline{\csname mod\x\endcsname}:~\csname modLong\x\endcsname,
	}
 	\underline{MMod}: Machine Model,
 	\underline{SEQ}: Sequential,
 	\underline{SHM}: Shared-Memory,
 	\underline{DM}: Distributed Memory,
 	\underline{Py}: Python
	}
  \label{tab:generators}
  \scalebox{0.89}{%
  	\small%
\begin{tabular}{l p{\softwareTableDescWidth}ll}
Toolkit   & Url           \& Models & Language & MMod \\
                  \midrule
                  \midrule
                  \multicolumn{4}{c}{Implementations of Multiple Models} \\
                  \midrule
                  \midrule

\software{GraphTool}
	{https://graph-tool.skewed.de}
	{\modES, \modRDT, \modSBM}
	{C++}{SHM}
\midrule

\software{GTGraph}
	{http://www.cse.psu.edu/~kxm85/software/GTgraph}
	{\modER, \modRMAT}
	{C}{SEQ}
\midrule

\software{IGraph}
	{https://igraph.org/}
	{\modBA, \modER,  \modES, \modSBM, \modWS}
	{C++, Py, R\iftrue\phantom{3}\fi}{SEQ}
\midrule

\software{KaGen}
	{https://github.com/sebalamm/KaGen}
	{\modBA, \modER, \modRDT, \modRGG, \modRHG}
	{C++}{SHM, DM}
\midrule

\software{NetworkX}
  {https://networkx.github.io/}
  {\modBA, Caveman, \modER, Holme-Kim, \modLFR, \modRGG,  \modSBM, \modWS}
  {Python}{SEQ}
\midrule

\software{NetworKit}
	{https://networkit.github.io/}
	{\modBA, \modChungLu, Clustered Random Graphs, \modER, \modFDSM, PubWeb, \modRHG, \modRMAT}
	{C++, Py}
	{SHM}
\midrule

\software{Snap}
	{https://snap.stanford.edu/snap}
	{\modBA, \modCM, Forest Fire, Multiplicative Attribute Graphs, \modNodeCopy, \modRMAT}
	{C++}{SHM}

\midrule
\midrule
\multicolumn{4}{c}{Implementations of a Single Model} \\
\midrule
\midrule

\software{Darwini}
	{https://issues.apache.org/jira/browse/GIRAPH-1043}
	{\modDarwini}
	{Java}{DM}
\midrule

\software{FEASTPACK}
	{https://www.sandia.gov/~tgkolda/feastpack/}
	{\modBTER}
	{MATLAB}
	{SEQ}
\midrule

\software{Graph500}
	{https://graph500.org/}
	{\modRMAT}
	{C}
	{DM}
\midrule

\software{HyperGen}
	{https://github.com/manpen/hypergen}
	{\modRHG}
	{C++}{SM}
\midrule

\software{LFR}
	{https://sites.google.com/site/andrealancichinetti/files}{}
	{C++}
	{SEQ}
\midrule

\software{MUSKETEER}
	{https://github.com/sashagutfraind/musketeer}
	{planar version: \url{https://github.com/isafro/Planar-MUSKETEER}}
	{Python}
	{SEQ}
\midrule

\software{R-MAT}
	{https://github.com/lorenzhs/rmat}
	{\modRMAT}
	{C++}
	{SHM}

\bottomrule
\end{tabular}}
\end{table}

In this section, we aim to give a short overview over publicly available software packages as well as implementations of single models.
In principle, we try to avoid historic generators that are not widely used anymore.
We focus on tools that either can generate a wide-range of models and for those we report all the models that we covered within this survey, or software that is specialized on a single model.
An overview can be found in \Cref{tab:generators}. 
\section{Future Challenges}
\label{sec:future-challanges}
\iffalse
   \newcommand{\mpar}[1]{\paragraph{#1.}}
\else
   \newcommand{\mpar}[1]{\paragraph*{#1}}
\fi

Generating graphs remains a widely open field for future research.
It is an interesting question to what extent the multitude of algorithms that we sketched in this survey can be improved further or how techniques outlined can be used to derive algorithms for new or other models not discussed here.
We believe that there are plenty of open~problems.

\mpar{Parallelism and Hardware Issues}
As current parallel machines are able to run billions of threads, scalable graph generation remains an open problem for many models.
This becomes even more pronounced for supercomputer systems with millions of processors that often are hierarchically organized (\eg in islands, racks, nodes, CPU sockets, cores, and threads).
These hierarchies and heterogeneity make the implementations highly complicated.
One way to tackle this problem may be to design more algorithms that are either communication-free or communication efficient.
Still even sequential algorithms are often hard to get scalable.
A quite obvious challenge is to make not-yet-scalable graph generators scalable, either by clever engineering or simplifying the model.
For example \moddKGraphs{} are an interesting model, but there is little known how difficult it is to generate them.
While scalable implementations for $d=2$ seem possible, $d = 3$ might be an interesting challenge.

\mpar{Numerical Stability}
Many generators (\eg based on hash functions) use fewer random bits or a smaller pseudo-random state than required to allow the creation of every possible instance.
While implications of this issue are well understood for a number of random combinatorial objects (\eg in case of random permutations~\cite{SALFI1974, Knu81}), it is an open question if graph properties of interest are unbiased in such cases.
We expect that using fewer random bits yields a trade off between coverage and efficiency---however, this needs thorough investigation from both the theory and practical side.
This is particularly consequential for hypothesis-testing scenarios, where a generator is to create an unbiased ensemble.

Another possible source of bias are numerical instabilities which typically occur during floating point operations.
While moderately sized instances can be sampled with default standard library arithmetic/special functions, it gets challenging for huge distributed instances.
A simple approach to overcome this issue would be to use arbitrary precision libraries which is often practically infeasible.
Alternatives include to explicitly manage the errors (\eg~\cite{DBLP:journals/corr/abs-1204-5834}), or to use efficient and exact sampling methods as demonstrated by~\cite{DBLP:conf/icalp/BringmannF13} for \modER and \modChungLu.

\mpar{Models vs.\ Applications}
There is a gap between research on scalable graph generation algorithms and the applications in which they are used.
Oftentimes, the domain-specific properties a model should reflect are highly confounded, poorly formalized, or not even fully understood.
In these cases, the models described here correspond to rather idealized situations in which many details have been stripped away.
While they may be sufficiently close to warrant benchmarking of algorithms, they are unlikely to be suitable for statistical modeling~\cite{goldenberg2010survey,snijders2011models}.

An example of more expressive models are \modLongERGM (\modERGM)~\cite{lusher2013ergms} common in social network research.
Given, say, graph statistics $s_1,\ldots,s_k$ and an associated vector $\theta=(\theta_i)_{i=1,\ldots,k}$ of parameters, a graph $G\in\gGn$ is assigned probability $P_\theta(G) = \frac{1}{Z(\theta)}\exp\left(\sum_{i=1}^k \theta_i\cdot s_i(G)\right)$ where $Z(\theta)=\sum_{H\in\gGn} \exp\left(\sum_{i=1}^k \theta_i\cdot s_i(H)\right)$ is the normalizing constant.
Statistics are chosen based on theories regarding micro-level mechanisms that may explain the formation of a network.
They typically range from the number of edges, to counts of various other kinds of small subgraphs.
The \modERGM with the edge count statistic $s_1(G)=m$ and parameter $\theta_1=\ln\frac{p}{1-p}$, for instance, is equivalent to $\Gnp$.

Because of the potential generality and complexity of distributions, sampling from models such as \modERGM is usually done using Markov Chain Monte Carlo algorithms.
These are of limited scalability, and can benefit from algorithmic contributions towards more efficient updates of statistics after each step of the Markov Chain.
Limited scalability is also an issue for parameter estimation, a common task in network modeling.
Here, new approaches seem to be necessary to make such models scale to larger graphs (see, \eg~\cite{chatterjee2013ergm,stivala2020ergm}).

Moreover, many application domains suppose inter-dependencies between graph structures and other attributes of nodes and edges, as well as multiple types of edges and the evolution of graphs over time.
The identification of recurring principles and algorithmic building blocks may be one of the most important challenges in this area.

\mpar{Libraries and Portability}
It would be highly helpful to have well maintained libraries that are able to work on different models of computation including GPU, HPC, and BigData-Tools such as MapReduce~\cite{DBLP:journals/cacm/DeanG08}, Spark~\cite{DBLP:journals/cacm/ZahariaXWDADMRV16}, or Thrill~\cite{DBLP:conf/bigdataconf/BingmannAJLNNSS16}.
In principle, given the same random seeds and parameters of the model, the libraries should be able to guarantee that the graph generated is the same independent on the underlying platform that is used.
Then, different researchers will be able to perform experiments on the same graphs on very large machines without the need to transfer and manage huge amounts of data.
Moreover, such libraries could include plug and play algorithms like dropping random edges, union, dynamization, et cetera.
The obvious way to deal with portability would be (de)serialization of data and writing to disk (which is often a bottleneck).
It remains to be answered what data structures are a good fit in this case, or how to partition the data in distributed generators. 

\indexAliasesOfModels%
\indexAliasesOfAlgorithms%
\indexAliasesOfImplementations%
   {
      \small
      \bibliographystyle{plainurl}
      \bibliography{paper}
   }
\end{document}